\documentclass[prl,preprint,onecolumn,showpacs,floatfix,superscriptaddress]{revtex4-2}

\usepackage{graphicx}
\usepackage{subfigure}
\usepackage{bm}
\usepackage{amsmath}
\usepackage{tabularx}
\usepackage{array}
\usepackage{multirow}
\usepackage{float}
\usepackage{xcolor}
\usepackage{hyperref}
\usepackage{physics}
\usepackage{lineno}

\begin{document}
\title{Rapid and Unconditional Parametric Reset Protocol for Tunable Superconducting Qubits}
\author{Yu Zhou}
\thanks{These two authors contributed equally to this work.}
\affiliation{Tencent Quantum Laboratory, Tencent, Shenzhen, Guangdong 518057, China}
\author{Zhenxing Zhang}
\thanks{These two authors contributed equally to this work.}
\affiliation{Tencent Quantum Laboratory, Tencent, Shenzhen, Guangdong 518057, China}
\author{Zelong Yin}
\affiliation{Tencent Quantum Laboratory, Tencent, Shenzhen, Guangdong 518057, China}
\author{Sainan Huai}
\affiliation{Tencent Quantum Laboratory, Tencent, Shenzhen, Guangdong 518057, China}
\author{Xiu Gu}
\affiliation{Tencent Quantum Laboratory, Tencent, Shenzhen, Guangdong 518057, China}
\author{Xiong Xu}
\affiliation{Tencent Quantum Laboratory, Tencent, Shenzhen, Guangdong 518057, China}
\author{Jonathan Allcock}
\affiliation{Tencent Quantum Laboratory, Tencent, Shenzhen, Guangdong 518057, China}
\author{Fuming Liu}
\affiliation{Tencent Quantum Laboratory, Tencent, Shenzhen, Guangdong 518057, China}
\author{Guanglei Xi}
\affiliation{Tencent Quantum Laboratory, Tencent, Shenzhen, Guangdong 518057, China}
\author{Qiaonian Yu}
\affiliation{Tencent Quantum Laboratory, Tencent, Shenzhen, Guangdong 518057, China}
\author{Hualiang Zhang}
\affiliation{Tencent Quantum Laboratory, Tencent, Shenzhen, Guangdong 518057, China}
\author{Mengyu Zhang}
\affiliation{Tencent Quantum Laboratory, Tencent, Shenzhen, Guangdong 518057, China}
\author{Hekang Li}
\affiliation{Beijing National Laboratory for Condensed Matter Physics, Institute of Physics, Chinese Academy of Sciences, Beijing 100190, China}
\affiliation{School of Physical Sciences, University of Chinese Academy of Sciences, Beijing 100049, China}
\author{Xiaohui Song}
\affiliation{Beijing National Laboratory for Condensed Matter Physics, Institute of Physics, Chinese Academy of Sciences, Beijing 100190, China}
\affiliation{School of Physical Sciences, University of Chinese Academy of Sciences, Beijing 100049, China}
\author{Zhan Wang}
\affiliation{Beijing National Laboratory for Condensed Matter Physics, Institute of Physics, Chinese Academy of Sciences, Beijing 100190, China}
\affiliation{School of Physical Sciences, University of Chinese Academy of Sciences, Beijing 100049, China}
\author{Dongning Zheng}
\affiliation{Beijing National Laboratory for Condensed Matter Physics, Institute of Physics, Chinese Academy of Sciences, Beijing 100190, China}
\affiliation{School of Physical Sciences, University of Chinese Academy of Sciences, Beijing 100049, China}
\author{Shuoming An}
\email{shuomingan@tencent.com}
\affiliation{Tencent Quantum Laboratory, Tencent, Shenzhen, Guangdong 518057, China}
\author{Yarui Zheng}
\affiliation{Tencent Quantum Laboratory, Tencent, Shenzhen, Guangdong 518057, China}
\author{Shengyu Zhang}
\affiliation{Tencent Quantum Laboratory, Tencent, Shenzhen, Guangdong 518057, China}

\begin{abstract}

Qubit initialization is a critical task in quantum computation and communication. Extensive efforts have been made to achieve this with high speed, efficiency and scalability. However, previous approaches have either been measurement-based and required fast feedback, suffered from crosstalk or required sophisticated calibration. Here, we report a fast and high-fidelity reset scheme, avoiding the issues above without any additional chip architecture. By modulating the flux through a transmon qubit, we realize a swap between the qubit and its readout resonator that suppresses the excited state population to 0.08\% $\pm$ 0.08\% within 34 ns (284 ns if photon depletion of the resonator is required).
Furthermore, our approach (i) can achieve effective second excited state depletion, (ii) has negligible effects on neighbouring qubits, and (iii) offers a way to entangle the qubit with an itinerant single photon, useful in quantum communication applications.

\end{abstract}
\maketitle

\section{Introduction}

Qubit initialization is fundamental and crucial for many quantum algorithms and quantum information processing tasks. The ability to quickly reset qubits to the zero state is one of DiVincenzo's essential criteria for building a quantum computer \cite{DiVincenzo2000} and is critical for quantum error correction \cite{schindler2011, fowler2012surface, reed2012realization}, where the reset of syndrome qubits needs to be accomplished with high fidelity in the time scale of a single qubit pulse. Furthermore, significant reduction of state preparation and measurement (SPAM) errors can be achieved by evacuating residual excited state populations with high fidelity \cite{slichter2012, riste2012}. The simplest way to reset qubits is to passively wait for them to de-excite, but as qubit relaxation times increase beyond 100 $\mu$s \cite{rigetti2012superconducting, place2020}, this method becomes impractically slow. Alternatively, active reset implementations can shorten the wait time between cycles and significantly improve computational efficiency \cite{egger2018, geerlings2013}.

Various reset protocols for superconducting qubits have been proposed which fall into two main types: measurement- and non-measurement-based protocols.
In measurement-based schemes, a qubit is measured and either heralded in the ground state \cite{johnson2012}, or else is found to be in the excited state and reset via a conditional $\pi$-pulse \cite{riste2012,riste2012b,salathe2018,campagne2013,corcoles2021exploiting}.
These protocols depend heavily on measurement fidelity and suffer from
measurement-induced state mixing \cite{slichter2012,roch2014}. In addition, the hardware implementation of necessary short-latency feedback loops is also a challenge.  In non-measurement based protocols, qubits are coupled to a lossy environment, usually a resonator. While numerous approaches to this have been proposed, they each suffer from their own drawbacks. For instance, in one such approach, flux control \cite{reed2010,mcewen2021removing} is used to rapidly tune the qubit frequency to that of the resonator. However, this process significantly affects neighbouring qubits via crosstalk \cite{krantz2019,kelly2014orbit}.
Another approach is based on a microwave-induced interaction between the qubit and a low-quality factor resonator \cite{magnard2018,egger2018}. However, the involvement of the second excited state $\ket{f}$ makes these schemes complicated and necessitates sophisticated calibration.
Furthermore, intense microwave driving is required to activate the required cavity-assisted Raman processes \cite{magnard2018, zeytinouglu2015, pechal2014}, affecting adjacent qubits as well.  In \cite{magnard2018}, an additional resonator is required to achieve the best performance.
In contrast to the above methods, the driven reset scheme proposed in \cite{geerlings2013} is free from flux control and complicated pulses. On the other hand, this protocol requires that the resonator dissipation rate $\kappa_r$ be smaller than the dispersive shift $\chi$, imposing a trade-off between readout speed and fidelity.

In this work, we demonstrate a rapid and unconditional parametric reset scheme for tunable superconducting qubits.
By parametric modulation of the qubit frequency, a controllable interaction is generated between the qubit and a lossy readout resonator. This interaction unconditionally transfers the qubit excitation to the resonator and thus resets the qubit on demand.
Using this method, we can suppress the residual excited population to 0.08\% $\pm$ 0.08\% within 34 ns.
We also demonstrate effective $\ket{f}$ state depletion in the case when leakage to higher states is non-negligible. Our protocol only involves AC modulation of at most two frequencies and does not need sophisticated calibration. Moreover, it has a negligible effect on subsequent gates and other qubits. It is compatible with circuit quantum electrodynamics systems \cite{blais2004,wallraff2004nature,blais2020cqedreview} and can be applied to all frequency-tunable superconducting qubits, requiring no additional hardware or modifications to chip components.
The method also imposes no restriction on operation flux position or specific system parameters such as resonator dissipation rate $\kappa_r$ or dispersive shift.

\section{Results}
\begin{figure}
\centering
\includegraphics[width=0.8\linewidth]{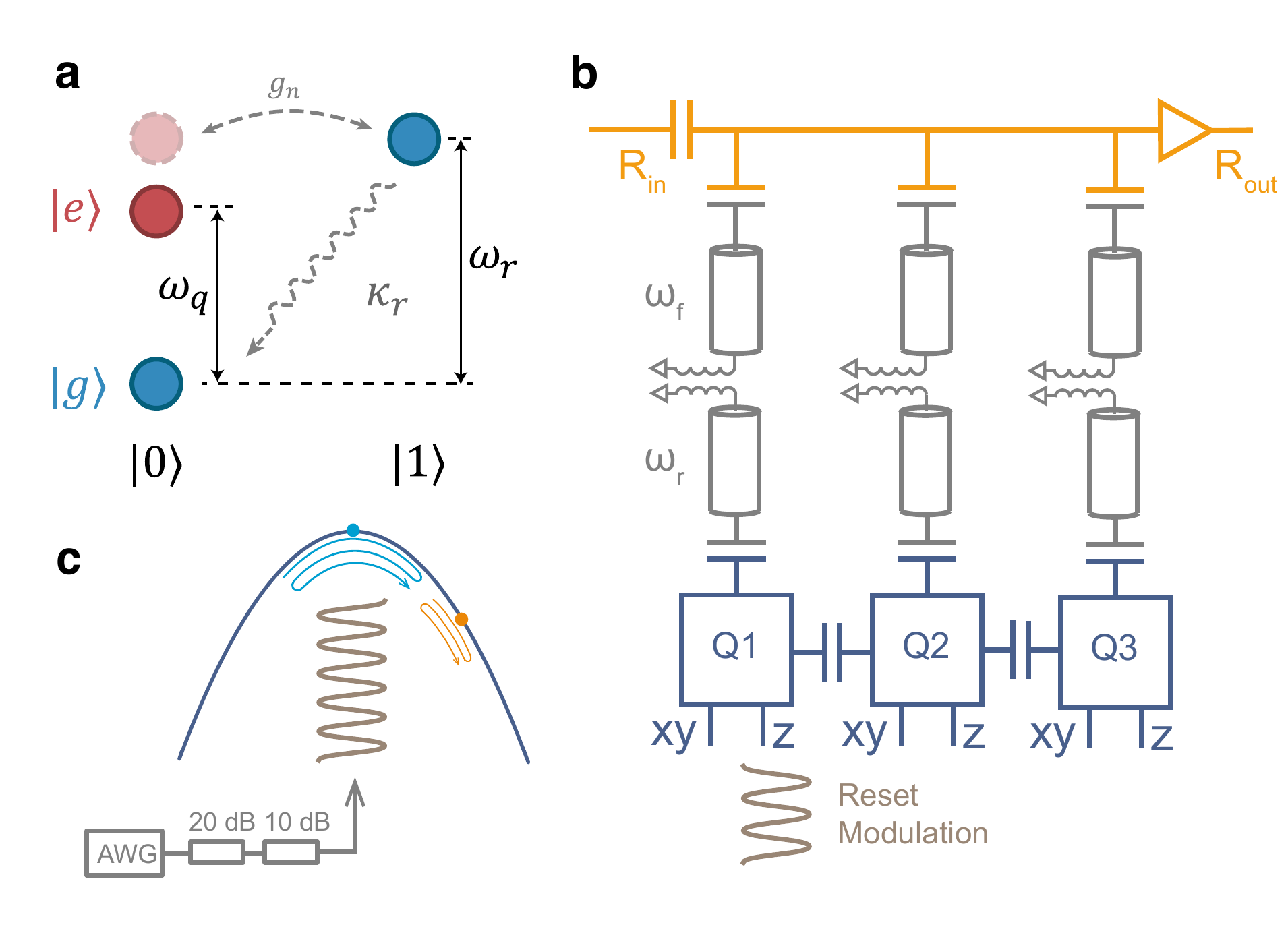}
\caption{ \textbf{Schematic of the reset process and diagram of the device. }
\textbf{a} Jaynes-Cummings ladder diagram of the qubit-resonator coupled system.  $\ket{g}$ , $\ket{e}$ and $\ket{0}$, $\ket{1}$ denote the qubit and resonator lowest two states respectively. The dashed light red circle represents one of the sideband modes induced by the parametric modulation. The dashed arrow labelled $\kappa_r$ illustrates the decay process of the resonator. \textbf{b} Simplified circuit diagram of the chip. Each transmon qubit is coupled to a readout resonator and an individual Purcell filter. The readout signal is amplified by an impedance-matched Josephson parametric amplifier (IMPA).
\textbf{c} Typical transmon resonance frequency $\omega_{\mathrm{q}}$ with respect to the flux
bias. The reset pulse is generated by AWG. After 30 dB attenuation,
it is added to the flux control line. The qubit's frequency modulation (brown) activates effective coupling between the qubit and its readout resonator and facilitates the reset process. Two cases are depicted: (1) operation point at or near the sweet spot (cyan, main text). (2) operation point away from the sweet spot (orange, Supplementary Note 5).}
\label{fig:figure1}
\end{figure}

\textbf{Theory. }
Our qubit reset protocol is based on a parametric activated interaction between
a tunable qubit and a rapidly decaying resonator.
Such a parametric modulation induces an effective tunable coupling between the qubit and other quantum systems such as another qubit or resonator \cite{li2013motional,beaudoin2012,strand2013} and has been used to implement multi-qubit quantum gates \cite{reagor2018,didier2018,caldwell2018,chu2020,mckay2016,hong2020}, state transfer \cite{li2018,luyao2017}, switches for quantum circuits \cite{wuyulin2018switch} and parity measurements \cite{royer2018}.
In our reset protocol, the parametric modulation induces Rabi oscillations between $\ket{e,0}$ and $\ket{g, 1}$,
where $\ket{s, l}$ denotes the tensor product of the qubit state $\ket{s}$ (the cases $\ket{s} = \ket{g}$ and $\ket{s} = \ket{e}$ correspond to the ground and excited states, respectively) and the resonator Fock state $\ket{l}$.
When the qubit is excited,  as illustrated in Fig.~\ref{fig:figure1}a,
the population can be transferred from
the qubit ($\ket{e,0}$) to the resonator ($\ket{g,1}$), which then rapidly decays to the target state $\ket{g,0}$ at decay rate $\kappa_r$,  which is mainly due to the large photon emission rate of the readout resonator.

We consider a qubit-resonator coupled system described by the Jaynes-Cummings model.
In the dispersive regime, there is no population exchange
due to the large detuning between the qubit and the resonator.
The external flux $\Phi$ is modulated as
$\Phi(t)=\overline{\Phi} + \Phi_m \cos(\omega_m t +\theta_m)$, where $\overline{\Phi}$ is the parking flux and $\Phi_m, \omega_m, \theta_m$ is the flux modulation amplitude, frequency and phase, respectively.
Due to the nonlinear dependence of the qubit frequency on the flux bias,
the qubit frequency $\omega_q(t)$ is, in general, described by a Fourier series with non-trivial higher-order terms, i.e.
$\omega_q(t) = \overline{\omega_q} + \sum_{k=1} A^{(k)}_m \cos[k( \omega_m t + \theta_m)]$
where $A_m^{(k)}$ are the Fourier coefficients
and $\overline{\omega_q}$ is the average frequency in the presence of the modulation~\cite{didier2018}.
In the case of small modulation, we take the leading term of
the qubit frequency as an approximation, i.e. $\omega_q(t) \approx \overline{\omega_q} + A_m^{(\alpha)} \cos[\alpha(\omega_m t +\theta_m)]$,
where $\alpha=1$ for the qubit parked away from the sweet spot, and $\alpha=2$ for the qubit parked in the sweet spot (in the latter case the odd Fourier coefficients $A^{(2k+1)}_m$ vanish~\cite{didier2018}).
The oscillation of the qubit frequency induces a series of
sidebands $\overline{\omega_q}+n\omega_m$, where $n$ is an integer.
When the frequency of one sideband satisfies the constraints
$n\omega_m = -\overline{\Delta} = \omega_r - \overline{\omega_q}$,
the transition between the states $\ket{e,l}$ and $\ket{g, l+1}$ is activated.
The effective coupling strength can be derived as
$g_n=\overline{g_{qr}}J_n(\frac{A^{(\alpha)}_m}{\omega_m})e^{i\beta_n}$,  where $\overline{g_{qr}}$ is the
averaged coupling strength between the qubit and the resonator during the modulation,
$J_n(x)$ are Bessel functions of the first kind, and $\beta_n=n\theta_m - \frac{A^{(\alpha)}_m}{\alpha\omega_m}\sin(\alpha\theta_m)$ is the interaction phase~\cite{didier2018}.

We consider the single excitation subspace spanned by $\{\ket{e,0}, \ket{g, 1}\}$, within which the dynamics of the reset
protocol can be modeled by the non-Hermitian Hamiltonian
\begin{equation}\label{eq:Heff}
H_\mathrm{eff} =
\begin{bmatrix}
0 & |g_n| e^{i\beta_n} \\ |g_n| e^{-i\beta_n} & -i \kappa_r/2
\end{bmatrix},
\end{equation}
where $|g_n|$ is the absolute value of $g_n$, and the non-Hermitian term $-i\kappa_r/2$ accounts for the decay of the photon in the resonator.
The population evolution can be expressed as
$P_{s|s_0}(t) = |\bra{s} e^{-iH_\mathrm{eff}t} \ket{s_0}|^2$,
where the system is initially prepared in the state $\ket{s_0}$,
and $\ket{s}$ is one of the states $\{\ket{e,0}, \ket{g,1}\}$.

The real parts of the eigenvalues $\{\lambda_k\}$ of $H_\mathrm{eff}$ determine the oscillation rate of $P_{s|s_0}(t)$, while the imaginary parts of $\{\lambda_k\}$ determine the exponential decay rates. We define the reset rate $\Gamma=2\min_k(|\mathrm{Im}[\lambda_k]|)$ as it is the smallest value of the decay rates and determines the overall protocol reset speed. Three different regimes are possible -- corresponding to overdamped, critically damped and underdamped oscillations of $P_{s|s_0}(t)$, respectively -- and our qubit reset works in all three regimes.
For small modulation amplitudes, i.e. $|g_n| < \kappa_r/4$, the reset is in the overdamped regime where the excited state population decays without oscillating.
In this regime, the reset rate $\Gamma$ increases with the modulation amplitude. At the critically damped point $|g_n|=\kappa_r/4$, the population shows a maximum reset rate $\kappa_r/2$ with no oscillation.
When the modulation amplitude satisfies $|g_n| >\kappa_r/4$ the reset becomes underdamped, and the population oscillates at rate
$\sqrt{4|g_n|^2-\kappa_r^2/4}$, and the reset rate remains at $\kappa_r/2$.

\textbf{Experimental realization.}
Our experimental setup is depicted in Fig.~\ref{fig:figure1}b and consists of three transmon qubits \cite{koch2007, barends2013}. Each transmon is capacitively coupled to a resonator with frequency $\omega_{r}/2\pi$ from 6.44~GHz to 6.68~GHz and coupling strength $g_{qr}/2\pi$ around 80~MHz.
Individual Purcell filters \cite{heinsoo2018,bultink2020}, implemented by $\lambda/4$ resonators, are inductively coupled to each readout resonator, and XY control and flux control (Z) lines are coupled to each qubit.
Fig.~\ref{fig:figure1}c displays the frequency $\omega_q$ of a transmon qubit
with respect to the flux. The reset pulse is generated from an arbitrary waveform
generator (AWG). After 30 dB attenuation, it is fed into the Z line, which results in frequency modulation of the qubit.

\begin{figure}
\centering
\includegraphics[width=0.8\linewidth]{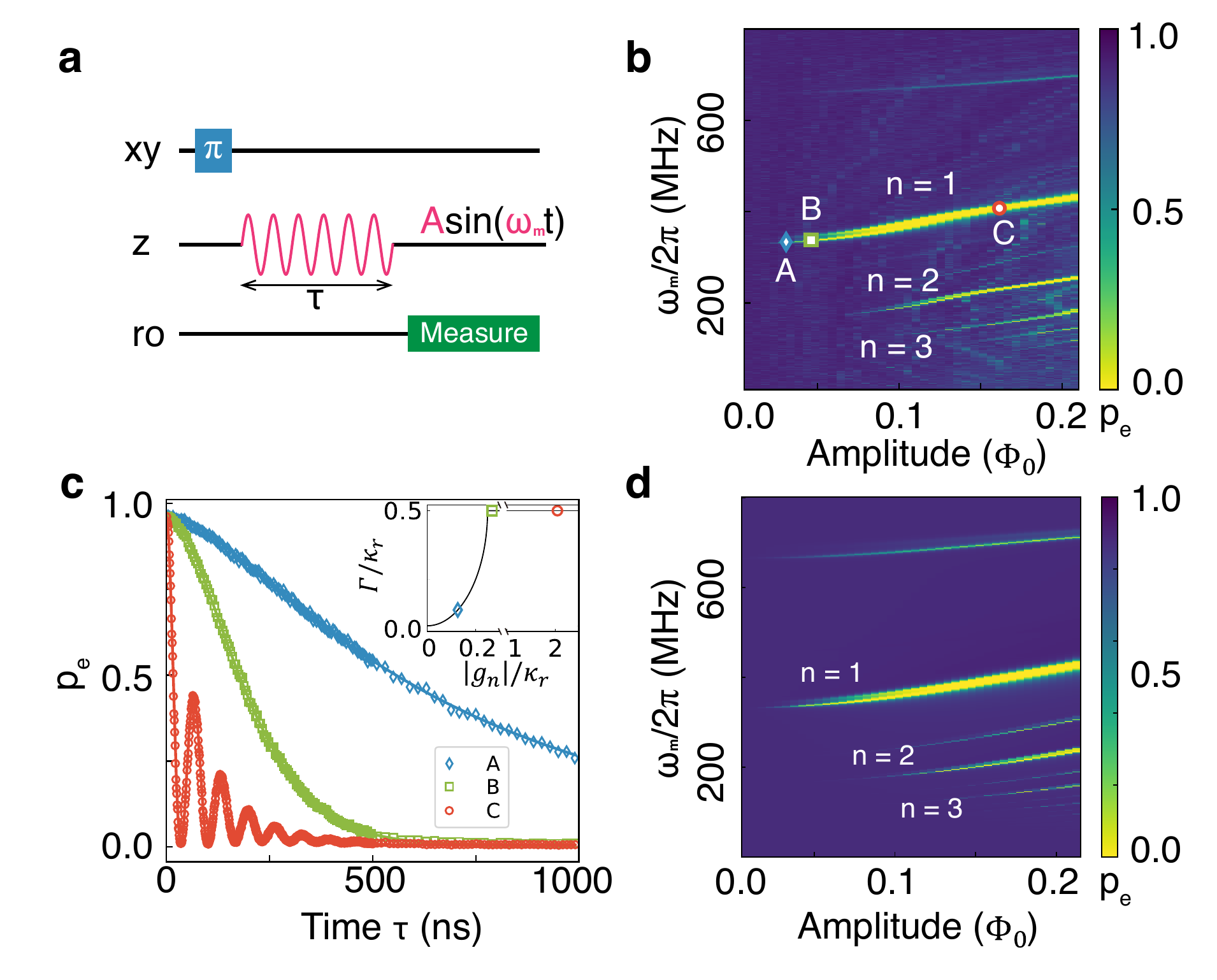}
\caption{\textbf{Realization of the parametric reset on Q1.}
\textbf{a} Experimental sequence of the parametric reset. A $\pi$-pulse is applied to the qubit and followed by a sinusoidal parametric reset pulse.
The amplitude $A$ and angular frequency $\omega_{m}$ are two adjustable parameters. \textbf{b} Two dimensional scan of the $\ket{e}$ population, $p_e$ when $\omega_{q}/2\pi = 5.784\, \mathrm{GHz}$, $\omega_{r}/2\pi = 6.441\, \mathrm{GHz}$, $\Delta/2\pi = -0.659\, \mathrm{GHz}$.
The x-axis is the parametric amplitude in the magnetic flux quantum $\Phi_0$
and the y-axis is the modulation frequency.
First, second, and third-order modulations are labelled $n=1,2,3$, respectively.
\textbf{c} Time evolution of the excited state population, after parametric modulation corresponding to the three points A, B, C in the $n=1$ region in (\textbf{b}).
Dots are the raw data acquired by direct readout measurements, and solid lines are fits to the theoretical model.
The insert shows theoretical values of $\Gamma/\kappa_r$ vs. $|g_n|/\kappa$, where $\Gamma$ is the reset rate of the qubit during the reset process.
\textbf{d} Master equation simulation of the whole process with the same parameters
as the experimental (\textbf{b}).}
\label{fig:figure2}
\end{figure}

We first demonstrate parametric reset on isolated Q1 (Q2, Q3 are tuned to their minimal frequencies through a fixed DC bias).
Fig~\ref{fig:figure2}a shows the detailed sequence:
A $\pi$-pulse is applied through the XY driveline to prepare the qubit in state $\ket{e}$.
A sinusoidal parametric reset pulse $A\sin(\omega_m t)$ of duration 1000 ns is applied
through the Z line, with amplitude $A$ and frequency $\omega_m$. Finally, Q1 is measured by the traditional dispersive readout. Fig.~\ref{fig:figure2}b shows the measured $\ket{e}$ population after this reset process, as a function of modulation amplitude $A$ (displayed in units of the magnetic flux quantum $\Phi_0$).

Several strips --- labeled $n = 1,2,3$, and corresponding to $n$-th order modulations --- are visible, where the population of $\ket{e}$ drops dramatically compared to other regions.
In these regions, 
one of the qubit's modulation sidebands is close to the resonator frequency, and the population transfers from the qubit to the resonator. When the operation point ($\omega_{q}/2\pi = 5.784\, \mathrm{GHz}$, $\omega_{r}/2\pi = 6.441\, \mathrm{GHz}$, $\Delta/2\pi = -0.659\, \mathrm{GHz}$) of the transmon qubit is close to the sweet spot, the qubit frequency will undergo two oscillations for every cycle of the parametric drive. As previously described in the theory section, the actual qubit modulation frequency is thus 2$\omega_m$, twice that of the flux modulation frequency $\omega_m$. There are several thin and unmarked strip-shaped regions in the figure due to the imperfect match between the operation point (0.004 $\Phi_0$) and the sweet spot. It is worth noting that sweet spot operation is not a requirement for our parametric reset protocol, and non-sweet spot operation is also suitable (see Supplementary Note 5).
Three points A, B, C in the $n=1$ region were selected and, for each point,
the qubit was first prepared in the $\ket{e}$ state and the population $p_e$ of $\ket{e}$ was then measured as a function of the duration of the parametric reset pulse $\tau$ by direct readout measurements (see Fig.~\ref{fig:figure2}c).
Corresponding to small modulation amplitude, point A (blue) lies in the overdamped regime where the $\ket{e}$ state population decays slowly and without any oscillation.  At point C (red), the modulation amplitude is large, and the $\ket{e}$ state population oscillates heavily, corresponding to the underdamped regime. Solid lines are fit to the theoretical model for $P_{s|s_0}(t)$ (Supplementary Note 4).
From the fitting, we extract $\kappa_r^{-1}$ of 46 ns, which agrees with
the direct measurement of the photon decay ($\kappa_r^{-1} \approx 50$ ns) of the resonator via AC Stark shift \cite{jeffrey2014}.
The insert displays
$\Gamma/\kappa_r$ vs. $|g_n|/\kappa_r$ predicted by theory, where $\Gamma$
is the reset rate for qubit state $\ket{e}$.
$\Gamma/\kappa_r$ increases in the overdamped regime and saturates at $\kappa_r/2$ in the underdamped regime.
Results for point B, chosen to be close to the critically damped point, are displayed in green in
Fig.~\ref{fig:figure2}c and show a population
decay much faster than in the overdamped regime, with no oscillations observed.
A master equation simulation was performed of the whole process using all experimental parameters, and the results are shown in Fig.~\ref{fig:figure2}d. Due to the limited sampling rate of the AWG,
the amplitude of the modulation signal is heavily attenuated by the analogue reconstruction filter at a high frequency. We reproduce this lowpass filter effect of the AWG in our simulation and find the main features agree with the experiment very well. In situations where high-frequency modulation is a must, (e.g. in the case of large detuning between the qubit and resonator and where the first-order region is preferred), the AWG can be replaced with a microwave source to overcome this limitation.

\begin{figure}
\centering
\includegraphics[width=0.8\linewidth]{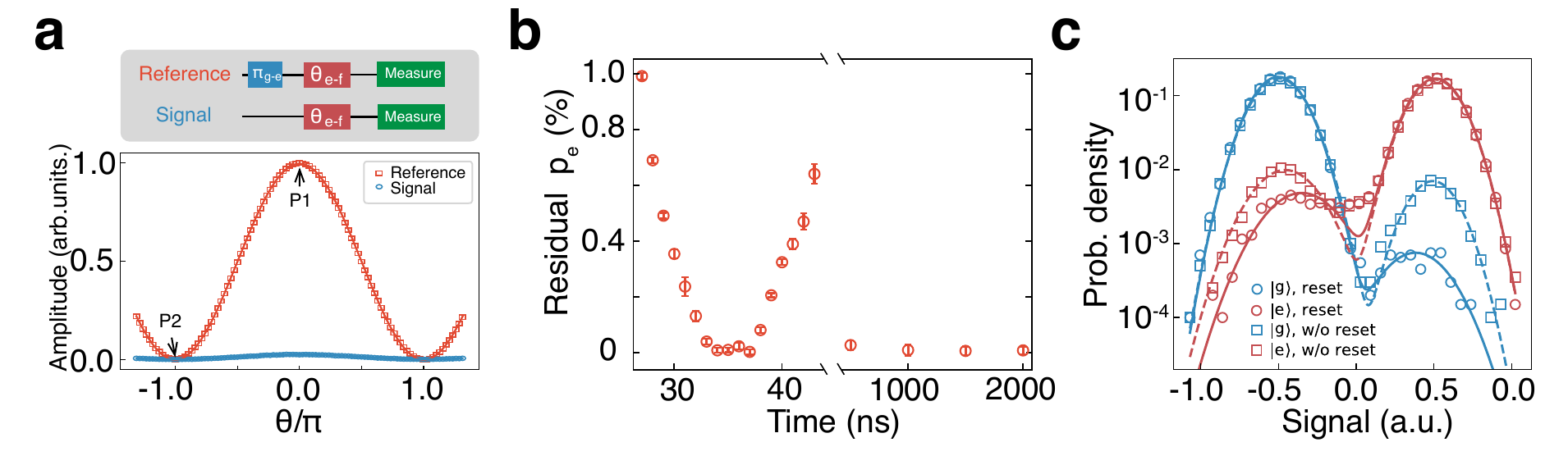}
\caption{\textbf{Residual $\ket{e}$ state population measurement and improvement of the
readout fidelity.} \textbf{a} Residual $\ket{e}$ population measurement with Reference (red)
and Signal (blue) Rabi oscillations with solid sinusoidal fitting curves. From the
fitting, two Rabi amplitudes($A_\mathrm{sig}$ and $A_\mathrm{ref}$) can be extracted.
The Reference and Signal pulse sequences are shown in the upper left corner. P1 and P2 are two
measured points in  (\textbf{b}). \textbf{b} $\ket{e}$ population measurement with different
parametric reset pulse duration $\tau$. Each data point is acquired by a
``two-point method" described in the main text. At 34 ns, the population of the
$\ket{e}$ decays to the first minimum value $0.08\pm0.08\%$ and remains below 0.1\%
after 1000 ns. The error bars are statistical ($\pm$1 s.d.) with 50 repetitions. \textbf{c} Readout fidelity enhancement with the 34 ns parametric reset
pulse. The circles (squares) display the IQ analysis with (without) the 34 ns
parametric pulse. Both $\ket{g}$ (blue) and $\ket{e}$ (red) are prepared and measured.
When preparing the $\ket{g}$ state, the residual $\ket{e}$ state
population due to the thermal excitation decreases more than one order of magnitude
after applying the parametric reset pulse. The readout fidelity consequently improves from 96.13$\%$ to 99.43$\%$ for $\ket{g}$ and 92.69$\%$ to 96.05$\%$ for $\ket{e}$ respectively.}
\label{fig:figure3}
\end{figure}

To measure the residual $\ket{e}$ state population, we perform a Rabi population
measurement (RPM) \cite{jin2015,geerlings2013} involving both the $\ket{e}$
and $\ket{f}$ (second excited) states in the sequence shown in Fig.~\ref{fig:figure3}a.
To acquire the Reference data (red squares), we first apply a $\pi$-pulse
to flip the population between the $\ket{g}$ and $\ket{e}$ states, then
perform a rotation around $X$ between $\ket{e}$ and $\ket{f}$ of angle $\theta$ ($\theta$-pulse).
Varying the angle $\theta$ results in Rabi oscillations of the $\ket{e}$
state population, as the Reference (red) shows. The solid line is a sinusoidal fitting. For the Signal data (blue circles), no $\pi$-pulse between $\ket{g}$ and $\ket{e}$ is applied.
However, there is still a visible and relatively small Rabi oscillation due to the residual
thermal $\ket{e}$ population. The final portion of residual $\ket{e}$
population $p_e$ is calculated by $A_\mathrm{sig}/(A_\mathrm{sig}+A_\mathrm{ref}$),
where $A_\mathrm{ref}$ and $A_\mathrm{sig}$ are the fitted amplitudes of the red
and blue oscillations, respectively. Without any parametric reset,
the measured $\ket{e}$ residual excited state population is around $2.38\pm0.06\%$
which corresponds to a 75 mK effective temperature.

We attribute this relatively high effective temperature to stray infrared radiation and insufficient thermalization of the sample box to the cold finger, which can be improved by careful shielding and better thermal contact~\cite{jin2015,kreikebaumSST16}.
The time evolution of point C is shown in Fig.~\ref{fig:figure2}c. When $\tau=34$ ns, the $\ket{e}$ state population decays
to its first minimum and reaches a steady low level after 1000 ns. Both the first minimum and steady-state points are of practical significance.
The former is useful when reset speed is the priority, and the reset protocol does not need to be reused immediately;  the latter is insensitive to parametric modulation time and requires fewer calibrations.  To reduce measurement error and accurately deduce the $\ket{e}$ residual population after the parametric reset pulse, a ``two-point method" \cite{jin2015} was used to increase the data acquisition efficiency.
Instead of measuring the whole trace as in Fig.~\ref{fig:figure3}a,
only the maximum and minimum points of the oscillation -- marked as P1, P2 -- were measured ($2\times10^5$ times each) to determine each value of $p_e$.
Fig.~\ref{fig:figure3}b shows the residual population of excited state $\ket{e}$ after variable parametric reset duration $\tau$ deduced via this two-point method. Each point corresponds to 50 measurements of $p_e$ with one standard deviation error bar. The minimum residual population reaches $0.08\pm0.08\%$ at 34 ns and remains below 0.1\% after 1000 ns, outperforming all existing reset schemes (Supplementary Note 3). Based on the rate model from Supplementary Note 6, we estimate the residual excitation population to be $0.02\%$ in thermal equilibrium.

The parametric reset process decreases state preparation error, yielding better
readout fidelity, as it effectively reduces the thermal population as illustrated by
Fig.~\ref{fig:figure3}c, where circles (squares) represents the
measurement with (without) the 34 ns parametric reset pulse.
With the parametric reset, the thermal excitation $p_e$ is reduced by more than an order of magnitude when preparing the $\ket{g}$ state.
Performing state discrimination analysis yields
a significant improvement of the readout fidelity with $\ket{g}$ from
96.13$\%$ to 99.43$\%$ and $\ket{e}$ from 92.69$\%$ to 96.05$\%$, respectively.

\begin{figure}
\centering
\includegraphics[width=0.8\linewidth]{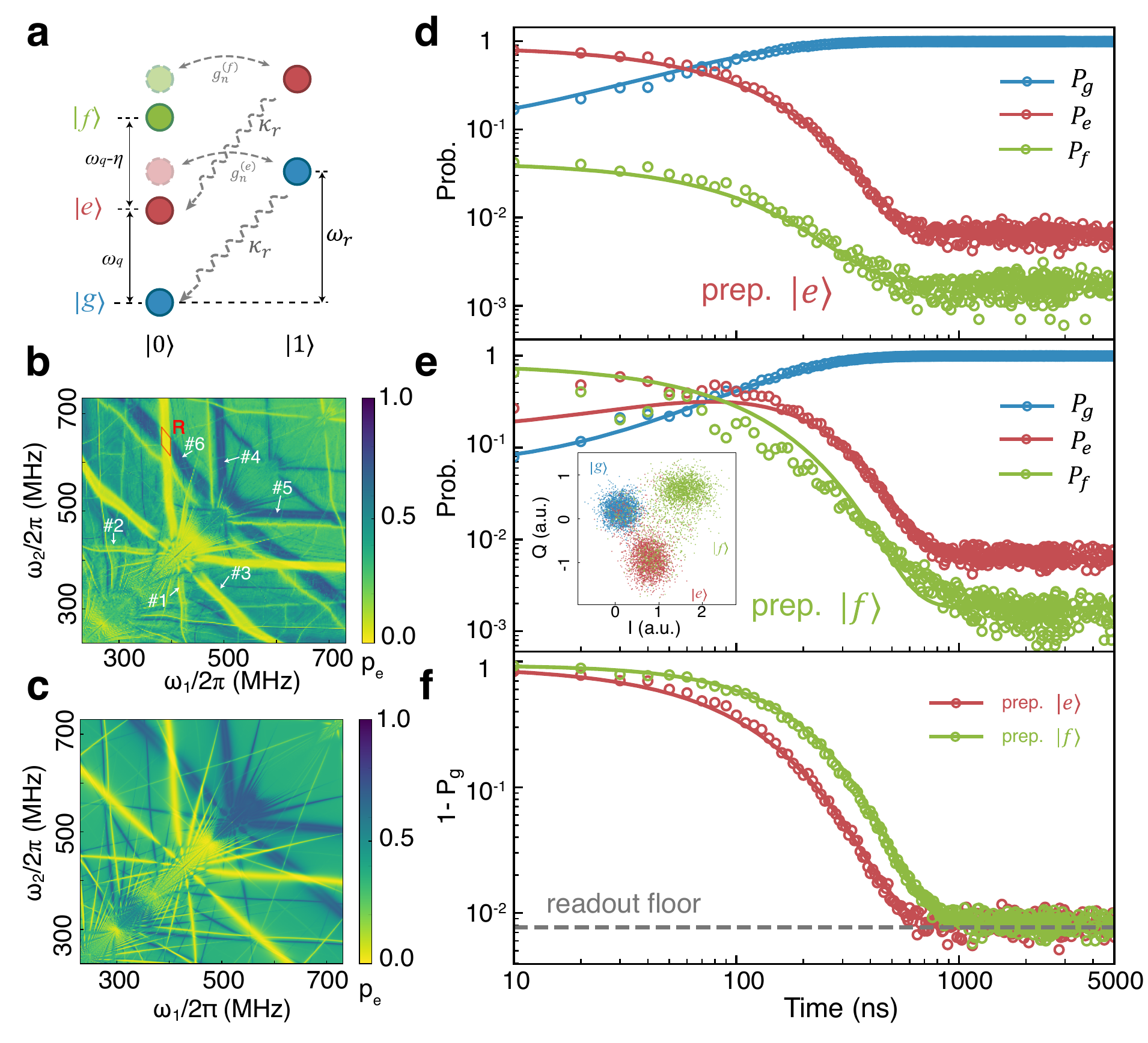}
\caption{\textbf{Two-tone parametric modulation.}
\textbf{a} Jaynes-Cummings ladder diagram of the two-tone parametric modulation process. $\ket{g}$, $\ket{e}$, $\ket{f}$ and $\ket{0}$, $\ket{1}$ denote the lowest energy states of the qubit and resonator, respectively. The dashed pale red and green circles represent one of the sideband modes induced by the parametric modulation. The dashed arrow labelled $\kappa_r$ illustrates the decay process of the resonator. \textbf{b} Scan map of the $\ket{e}$ population after 1000 ns two-tone  parametric modulation. Before modulation, the qubit was prepared in the $\ket{f}$ state. Blue and yellow strips correspond to the $\ket{f,0}\rightarrow \ket{e,1}\rightarrow\ket{e,0}$ and $\ket{e,0}\rightarrow\ket{g,1}\rightarrow\ket{g,0}$ decay processes, respectively.
\textbf{c} Master equation simulation with experimental parameters from \textbf{b}.
Time evolution of each qubit state, with the qubit initially prepared as $\ket{e}$ (\textbf{d}) or $\ket{f}$ (\textbf{e}), and then reset with two-tone modulation.
Inset figure in \textbf{e} is the IQ plot of each state.
\textbf{f} Time evolution of all remaining excited states ($1-P_g$). The grey dashed line is the readout floor.}
\label{fig:figure4}
\end{figure}

\textbf{Reset of the $\ket{f}$ state by two-tone parametric drive.}
Leakage to the second excited state $\ket{f}$ can be an important source of error during two-qubit gates~\cite{marques2021logical} and measurements~\cite{SankPRL16}.
In this section, we extend the single-tone parametric modulation scheme to one that uses two-tones in order to achieve effective $\ket{f}$ state reset.
In this case, the reset pulse has the form
$A_1\sin(\omega_1 t) + A_2\sin(\omega_2 t)$, where $A_1,A_2$ and $\omega_1,\omega_2$  are the amplitudes and frequencies of the two tones used, respectively, and the corresponding Fourier expansion of the qubit frequency has four main frequency components: $2\omega_1,2\omega_2,\omega_1\pm\omega_2$.
As shown in Fig.~\ref{fig:figure4}a, depletion of the $\ket{f}$ state comprises two processes. First, one frequency component of the parametric modulation causes $\ket{f,0}$ to interact with $\ket{e,1}$, with the latter decaying to $\ket{e,0}$ at the resonator dissipation rate $\kappa_r$. Similarly, $\ket{e,0}$ decays to $\ket{g,0}$ via the second frequency component of the modulation.
Fig.~\ref{fig:figure4}b is a scanned map of the $\ket{e}$ state population after a 1000 ns two-tone parametric reset pulse, with the qubit initially prepared in the $\ket{f}$ state.
We consider the case where $A_1 = A_2$, and scan $\omega_1/2\pi$ and $\omega_2/2\pi$ from 230 MHz to 730 MHz.
Two spider-like regions can be seen in the figure, one consisting of multiple blue strips and the other consisting of multiple yellow strips. These two regions correspond respectively to the first and second decay processes mentioned above.
Six of these colored strips are annotated in the figure,
where strips 1,2 (4,5) correspond to the regions where $2\omega_{1(2)} = -\overline{\Delta}$ ($2\omega_{1(2)} = -\overline{\Delta}-\eta$), and strip 3 (6) corresponds to the region where $\omega_1+\omega_2 = -\overline{\Delta}$ ($\omega_1+\omega_2 = -\overline{\Delta}-\eta$).
Here, $\eta = -254$ MHz is the anharmonicity of Q1.
We perform a master equation simulation of this two-tone parametric reset process and find that the results (Fig.~\ref{fig:figure4}c) closely agrees with the experiment. See Supplementary Note 8 for more scan maps and theoretical explanation.
In the rhombus area marked R in Fig.~\ref{fig:figure4}b, where strips 1 and 6 intersect, the two decay processes coexist, and the region is thus suitable for fast depletion of the $\ket{f}$ state.
In Fig.~\ref{fig:figure4}d-e we consider the case where $A_2 = 1.8A_1$ and measure the time evolution of the $\ket{g}$,$\ket{e}$,$\ket{f}$ states in the R region.
Circular data points in these figures are experimental data, and the solid lines fit to a multi-level decay model~\cite{peterer2015coherence}.
The in-phase(I) and quadrature(Q) components of each state from the readout are shown inset in Fig.~\ref{fig:figure4}e.
The qubit was prepared in the $\ket{e}$ (Fig.~\ref{fig:figure4}d) and $\ket{f}$ (Fig.~\ref{fig:figure4}e) states, respectively, and the population of all excited states $1-P_g$ is shown in Fig.~\ref{fig:figure4}f.
The excited population for both initial states shows nearly exponential decay and reaches the readout floor within 600 ns (initial state: $\ket{e}$) and 1000 ns (initial state: $\ket{f}$).
From the multi-level decay model~\cite{peterer2015coherence}, we estimate the decay rates to be $1/100$ ns for state $\ket{e}$ and $1/117$ ns for $\ket{f}$ during the reset process. The measured reset fidelity is 99.23$\%$, limited by the readout fidelity.

\begin{figure}
\centering
\includegraphics[width=0.8\linewidth]{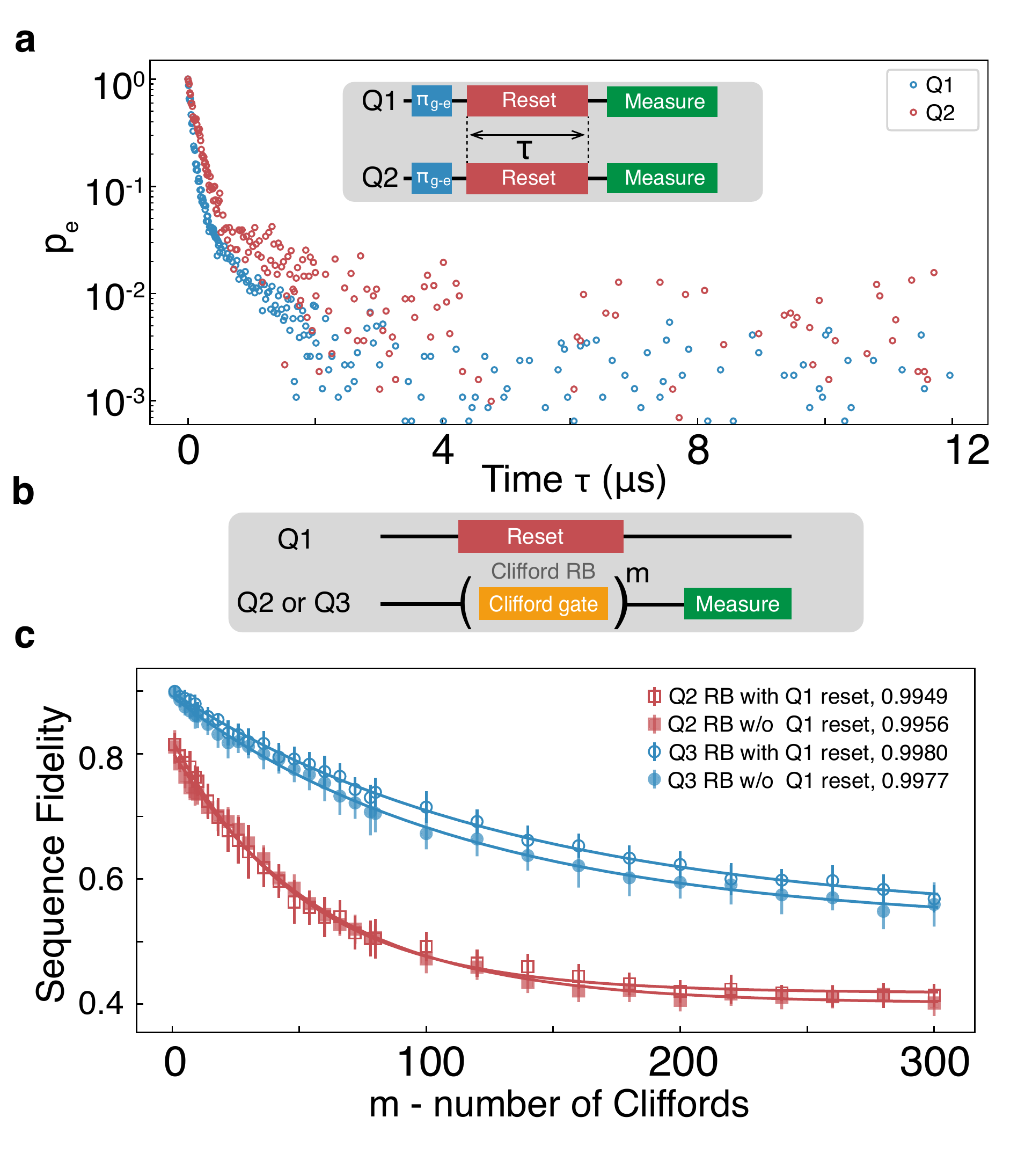}
\caption{\textbf{Simultaneous parametric reset of Q1, Q2, and effects of reset pulse on adjacent qubits.}
\textbf{a} Time evolution of Q1 (blue) and Q2 (red) $\ket{e}$ state population
when performing a parametric reset on both qubits simultaneously.
The population remains at a steady low level from 2 $\mu$s onwards. The pulse sequence for simultaneous parametric reset is inset.
\textbf{b} Pulse sequence for Clifford-based randomized benchmarking (RB) on Q2(Q3). During the RB, either a parametric reset (with the same pulse duration as the $m$ Clifford gates) was applied to Q1 or, for comparison, no pulse was applied to Q1. \textbf{c} RB results for the scenario described in \textbf{b}. The error bars are statistical ($\pm$1 s.d.) with 30 repetitions. The average single gate fidelity standard deviation is 0.08\%(0.03\%) for Q2(Q3). }
\label{fig:figure5}
\end{figure}

\textbf{Scalability of the protocol.}
To study the protocol's scalability, parametric resets were simultaneously performed on two qubits Q1 and Q2.
In this case, all qubits were tuned to operation points near their sweet spots.
The sequence is shown in the top part of Fig.~\ref{fig:figure5}a.
Both qubits are prepared in the $\ket{e}$ state by a $\pi$-pulse, and parametric modulation is applied separately to each qubit through their associated flux lines.
The time evolution of the excited state population $p_e$ of the qubits is measured with varied reset duration $\tau$. As seen in Fig.~\ref{fig:figure5}a, $p_e$ decays quickly and remains at a low level from 2 $\mu$s onwards, demonstrating the feasibility of our parametric reset protocol in a multi-qubit system.
One disadvantage of previous reset protocols involving flux pulses \cite{reed2010,mcewen2021removing} is that the Z pulse for the reset can significantly affect all subsequent gates \cite{foxen2018,rol2020cryoscope,barends2014,kelly2014orbit} or cause a frequency shift of neighbouring qubits \cite{zijunchen2018thesis}.
The parametric modulation we propose consists of one or two frequency components only and thus has negligible effects on neighbouring qubits. To prove this, Clifford-based randomized benchmarking (RB) was performed on Q2(Q3) when Q1 was reset with a single-tone parametric pulse (the results of RB with two-tone reset are given in Supplementary Note 10).
From the RB data (Fig.~\ref{fig:figure5}b), we find that the reset process decreases the average gate fidelity by only 0.08\% for nearest neighbour qubit Q2, and has almost no effect on the next-nearest neighbour qubit Q3, with only 0.03\% fidelity difference.
To further probe the effect of the reset process on the neighbouring qubits, we have also performed a series of Ramsey measurements, with results given in Supplementary Note 7.
Together, these experiments demonstrate that our parametric reset protocol has a negligible effect on adjacent qubits in terms of coherence, frequency and gate fidelity.

\section{Discussion}
In our single-tone parametric scheme, the first minimum point has practical significance when there is no immediate gate on the same qubit or when gates are immediately applied to other qubits, such as in state transfer \cite{kurpiers2018} or quantum simulations \cite{ma2019}. In scenarios where photon depletion of the resonator after qubit reset is a must, a ~5/$\kappa$ (250 ns) duration passive wait is sufficient to deplete the mean photon number below 0.01, corresponding to a negligible stark shift of 51.5 kHz. Taking re-thermalization (Supplementary Note 6) into consideration, the net fidelity at the end of this process is 99.86\%. An alternative way to achieve photon depletion of the resonator is to reset the resonator actively~\cite{mcclure2016rapid,bultink2016active}. If speed is not a priority, a one-tone reset with the modulation always on is an option, requiring 1000 ns and achieving 99.9\% fidelity. When $\ket{f}$ state leakage is not negligible, two-tone modulation is preferable and it takes 1000 (600) ns for both the $\ket{e}$ and $\ket{f}$ state ($\ket{e}$ only) to reach the 0.77\% readout floor. In conclusion, the parametric reset schemes we proposed have a high degree of flexibility that allows them to be used in a variety of different scenarios. We summarize the performance and use case scenarios of our protocols in Supplementary Table 2.

We have demonstrated a parametric reset protocol realized in transmon qubits which can be completed in 34 ns (284ns if one waits an additional five times the resonator $T_1^r = 1/\kappa_r$ time for the resonator to deplete). The speed and fidelity 99.92\% (99.86\% if ~5/$\kappa_r$ is included) of our approach outperforms all existing reset schemes (Supplementary Note 3) and, furthermore, has the added advantages of flexibility and scalability.
In theory, the reset time can be further decreased to less than 10 ns by increasing the modulation amplitude of the reset pulse or by increasing the coupling strength $g_{qr}$ in the qubit design. Moreover, as the RB and the Ramsey experiments show, our parametric modulation induces negligible effects on neighbouring qubits in terms of gate fidelity, frequency and coherence. By extending the method to using two-tone modulation, we are also able to achieve effective $\ket{f}$ state depletion. Our methods give a practical and universal way to reset tunable superconducting qubits and offer a pathway to achieving high-fidelity reset in large-scale qubit systems. Beyond qubit reset, parametric modulation-induced interaction can also be used in thermodynamic reservoir engineering \cite{luyao2017,kienzler2015} and quantum many-body simulations \cite{ma2019}. Furthermore, this work provides an efficient way to entangle the qubit state with an itinerant single photon, particularly useful in quantum communication and quantum network application \cite{kurpiers2018,kurpiers2019,ypzhong2020}.

\section{Methods}
\textbf{Hamiltonian with parametric modulation.}
We consider a qubit-resonator coupled system, which can be described
by the Jaynes-Cummings model ($\hbar=1$ hereafter)
\begin{equation}\label{eq:sys_ham}
H_\mathrm{sys} = \omega_q \ketbra{e}{e} + \omega_r a^\dag a
+ g_{qr} (a^\dag \sigma_- + a\sigma_+),
\end{equation}
where $\omega_q\; (\omega_r)$ is the qubit (resonator) frequency, $g_{qr}$ is the coupling strength between the qubit and the resonator and
$\sigma_+\; (\sigma_-)$ is the creation (annihilation) operator of the qubit.
In the interaction picture, when the qubit frequency is modulated as $\omega_q(t) \approx \overline{\omega_q} + A_m^{(\alpha)} \cos[
\alpha(\omega_m t + \theta_m)]$ ($\alpha$ an integer), to leading-term approximation
the system Hamiltonian can be expressed as
\begin{equation}\label{eq:sys_hamI}
H_\mathrm{int} = \sum_{n=-\infty}^{\infty} g_n e^{i(n\alpha\omega_m +\overline{\Delta})t} a\sigma_+ + h.c.
\end{equation}
where $g_n = \overline{g_{qr}} J_n\left(\frac{A_m^{(\alpha)}}{\alpha\omega_m}\right) e^{i\beta_n}$
are the effective coupling strengths in the leading term,  $\overline{g_{qr}}$ is the averaged coupling strength during the modulation,
$\beta_n = n\alpha\theta_m - \frac{A_m^{(\alpha)}}{\alpha\omega_m}\sin(\alpha\theta_m)$ is the interaction phase,
and $J_n(x)$ are Bessel functions of the first kind, and
$\overline{\Delta} = \overline{\omega_q} - \omega_r $ is the effective detuning between the qubit and the resonator during the sinusoidal modulation~\cite{didier2018}.
Note that for transmon qubits, the modulation of the qubit frequency also induces a modulation of the coupling strength, the full expression of which can be found in reference~\cite{didier2018}.

When the modulation frequency $\omega_m$ satisfies the constraint
$ n\alpha\omega_m + \overline{\Delta} = 0$ (for integer $n$) the Hamiltonian approximates to $H_\mathrm{int}=g_n a\sigma_+ + g_n^* a^\dag\sigma_-$ by ignoring rapidly oscillating terms, and Rabi oscillations occur between states $\ket{e, l}$ and $\ket{g, l+1}$.

\textbf{Qubit reset rate.}
When the qubit is prepared in the excited state $\ket{e}$,
the time dependent population can be solved by the effective
Hamiltonian $H_\mathrm{eff}$ of equation \eqref{eq:Heff}. The reset rate $\Gamma$ can be derived as:
\begin{equation}\label{eq:gamma}
\Gamma=2\min_{k}(|\mathrm{Im}[\lambda_k]|) =
\begin{cases}
\frac{1}{2}\left(\kappa_r - \sqrt{\kappa_r^2-16|g_n|^2}\right) & |g_n| < \kappa_r/4\\
\kappa_r/2  & |g_n| \geq \kappa_r/4\\
\end{cases}
\end{equation}
For $|g_n| \geq \kappa_r/4$, the reset rate remains at $\kappa_r/2$ and is independent of the effective coupling $|g_n|$.
For $|g_n| < \kappa_r/4 $, the reset rate $\Gamma$ increases with the effective coupling. In both cases, the reset rate is larger than the free decay rate of the qubit. The population of the qubit during the parametric reset process
can be modeled by
\begin{equation}\label{eq:popt_general}
P_{s|s_0}(t) = |\bra{s}\exp(-i H_\mathrm{eff} t)\ket{s_0}|^2,
\end{equation}
where the system is initially prepared in the state $\ket{s_0}$, and $\ket{s}\in\{\ket{e,0},\ket{g,1}\}$. When $\ket{s_0} = \ket{e,0}$,
the population of the excited state $p_e$ can be shown to be
\begin{equation}\label{eq:pe}
p_e = P_{e|e}(t)= \begin{cases}
e^{-\frac{\kappa_r t}{2}} \left(\frac{\kappa_r t}{4} + 1\right)^2 & |g_n| = \kappa_r/4\\
e^{-\frac{\kappa_r t}{2}} \left[\cos(Mt)+\frac{\kappa_r}{4M}\sin(Mt)\right]^2
& |g_n|>\kappa_r/4,\, M=\frac{\sqrt{16|g_n|^2-\kappa_r^2}}{4}\\
e^{-\frac{\kappa_r t}{2}} \left[\cosh(Mt)+\frac{\kappa_r}{4M}\sinh(Mt)\right]^2
& |g_n|<\kappa_r/4,\, M=\frac{\sqrt{\kappa_r^2 - 16|g_n|^2}}{4}\\
\end{cases}
\end{equation}

\textbf{Data availability.} Source data to generate figures and tables are available from the corresponding authors.

\section{Acknowledgments}
We acknowledge the support from the Key-Area Research and Development Program of Guangdong Province (2020B0303030002 and 2020B0303030001), the State Key Development Program for Basic Research of China (Grant No. 2017YFA0304300) and the Strategic Priority Research Program of Chinese Academy of Sciences (Grant No. XDB28000000).
\section{Author contributions}
S.M.A. and X.G. conceived the experiment and developed the theory. Y.Z., Z.X.Z. and S.N.H. built the set-up and carried out the measurement. Z.L.Y. and Z.X.Z. performed numerical simulations. Y.Z., Z.X.Z., S.M.A. and Y.R.Z. analyzed the results. Y.Z. and Z.X.Z. wrote the manuscript. J.A. and S.M.A. edited the manuscript. Z.X.Z. and X.X. developed the software platform for the experiment. H.K.L., X.H.S., Z.W. and D.N.Z. prepared the sample. F.M.L., G.L.X, Q.N.Y, M.Y.Z and H.L.Z. developed room-temperature electronics. S.Y.Z. supervised the project. All authors contributed to the discussion of the results and the development of the manuscript.
\section{Competing interests}
The authors declare no competing interests.

\end{document}


\title{Supplementary material for Rapid and Unconditional Parametric Reset for Tunable Superconducting Qubits}
\author{Yu Zhou}
\thanks{These two authors contributed equally to this work.}
\affiliation{Tencent Quantum Laboratory, Tencent, Shenzhen, Guangdong 518057, China}
\author{Zhenxing Zhang}
\thanks{These two authors contributed equally to this work.}
\affiliation{Tencent Quantum Laboratory, Tencent, Shenzhen, Guangdong 518057, China}
\author{Zelong Yin}
\affiliation{Tencent Quantum Laboratory, Tencent, Shenzhen, Guangdong 518057, China}
\author{Sainan Huai}
\affiliation{Tencent Quantum Laboratory, Tencent, Shenzhen, Guangdong 518057, China}
\author{Xiu Gu}
\affiliation{Tencent Quantum Laboratory, Tencent, Shenzhen, Guangdong 518057, China}
\author{Xiong Xu}
\affiliation{Tencent Quantum Laboratory, Tencent, Shenzhen, Guangdong 518057, China}
\author{Jonathan Allcock}
\affiliation{Tencent Quantum Laboratory, Tencent, Shenzhen, Guangdong 518057, China}
\author{Fuming Liu}
\affiliation{Tencent Quantum Laboratory, Tencent, Shenzhen, Guangdong 518057, China}
\author{Guanglei Xi}
\affiliation{Tencent Quantum Laboratory, Tencent, Shenzhen, Guangdong 518057, China}
\author{Qiaonian Yu}
\affiliation{Tencent Quantum Laboratory, Tencent, Shenzhen, Guangdong 518057, China}
\author{Hualiang Zhang}
\affiliation{Tencent Quantum Laboratory, Tencent, Shenzhen, Guangdong 518057, China}
\author{Mengyu Zhang}
\affiliation{Tencent Quantum Laboratory, Tencent, Shenzhen, Guangdong 518057, China}
\author{Hekang Li}
\affiliation{Beijing National Laboratory for Condensed Matter Physics, Institute of Physics, Chinese Academy of Sciences, Beijing 100190, China}
\affiliation{School of Physical Sciences, University of Chinese Academy of Sciences, Beijing 100049, China}
\author{Xiaohui Song}
\affiliation{Beijing National Laboratory for Condensed Matter Physics, Institute of Physics, Chinese Academy of Sciences, Beijing 100190, China}
\affiliation{School of Physical Sciences, University of Chinese Academy of Sciences, Beijing 100049, China}
\author{Zhan Wang}
\affiliation{Beijing National Laboratory for Condensed Matter Physics, Institute of Physics, Chinese Academy of Sciences, Beijing 100190, China}
\affiliation{School of Physical Sciences, University of Chinese Academy of Sciences, Beijing 100049, China}
\author{Dongning Zheng}
\affiliation{Beijing National Laboratory for Condensed Matter Physics, Institute of Physics, Chinese Academy of Sciences, Beijing 100190, China}
\affiliation{School of Physical Sciences, University of Chinese Academy of Sciences, Beijing 100049, China}
\author{Shuoming An}
\email{shuomingan@tencent.com}
\affiliation{Tencent Quantum Laboratory, Tencent, Shenzhen, Guangdong 518057, China}
\author{Yarui Zheng}
\affiliation{Tencent Quantum Laboratory, Tencent, Shenzhen, Guangdong 518057, China}
\author{Shengyu Zhang}
\affiliation{Tencent Quantum Laboratory, Tencent, Shenzhen, Guangdong 518057, China}

\maketitle


\section{Supplementary Note 1. Device Parameters}

\begin{table}[htb]
\centering
\begin{tabular}{ |p{3cm}||p{2.5cm}|p{2.5cm}|p{2.5cm}|  }
\hline
 & Q1 & Q2 & Q3\\
 \hline
 $\omega_{r}/2\pi$ (GHz)   & 6.441    &6.553&   6.686\\
 $\omega_{q}^{\mathrm{max}}/2\pi$ (GHz)   & 5.784    &5.393&   5.813\\
 $\omega_{q}^{\mathrm{idle}}/2\pi$ (GHz) &  5.783 ($5.780^{*}$)  & 5.005(($4.999^{*}$)   &5.686 ($5.811^{*}$)\\
 $T_1 \;(\mu \mathrm{s})$ &11.5 ($5.34^{*}$)& 11.4 ($5.66^{*}$)&  9.04 ($10.1^{*}$)\\
$T_2^{*} \;(\mu \mathrm{s})$    &11.0 ($13.6^{*}$) & 1.05 ($1.03^{*}$)&  0.89 ($25.14^{*}$)\\
$1/\kappa_r \;(\mathrm{ns}) $&   50  & 34&33\\
$g_{qr}/2\pi$ (MHz) & 78  & 83   &89\\
 \hline
$J$/2$\pi$ (MHz) & \multicolumn{3}{|c|}{Q1,Q2:17.5 \quad Q2,Q3:17.6} \\
 \hline
\end{tabular}
\caption{Device parameters. $\omega_r$, $\omega_q^\mathrm{max}$ and $\omega_q^\mathrm{idle}$ denote the frequency of the readout resonator, qubit frequency at the sweet spot, and operation frequency of the qubit, respectively. $g_\mathrm{qr}$ is the coupling between the qubit and the resonator. $J$ is the coupling strength between neighbouring qubits. * denotes the qubits' parameters in Supplementary Figure \ref{fig:figures6} b-e and Supplementary Figure \ref{fig:figures10}.}
\label{tab:params}
\end{table}

Our experiments are implemented on three transmon qubits -- Q1, Q2 and Q3 -- with parameters listed in Supplementary Table~\ref{tab:params}.
The resonator frequency $\omega_r/2\pi$ ranges from 6.441 to 6.686 GHz. The maximum frequency $\omega_q$/2$\pi$  of Q1 (5.784 GHz) is close to Q3 (5.813 GHz), and about 400 MHz higher than Q2 (5.393 GHz). Each resonator's $\kappa_r$ is measured individually via AC stark shift spectroscopy. Direct adjacent capacitive coupling is around 17.5 MHz.

\section{Supplementary Note 2: Experimental setup}
A schematic of our experimental setup is displayed in Supplementary Fig.~\ref{fig:figureS1}. The parametric modulation signal is generated by a home-made Arbitrary Waveform Generator (AWG) with sampling rate 2 Gs/s. After 30 dB attenuation, the signal is combined with a DC signal using a home-made bias tee (flux line). 
For XY and read-in lines, the baseband signal is generated by AWG, and then up-converted to the driving frequency by IQ mixer with a carrier LO signal generated by the microwave source.
The read-out signal from the qubit is first amplified by an impedance-transformed Josephson parametric amplifier(IMPA) \cite{mutus2014} and High Electron Mobility Transistors (HEMTs) at 4K.  
It is further amplified by room-temperature amplifiers and digitized by an analog-to-digital converter (ADC), before being demodulated and analyzed by DAQ FPGA.

\begin{figure}[htb]
\centering
\includegraphics[width=0.8\linewidth]{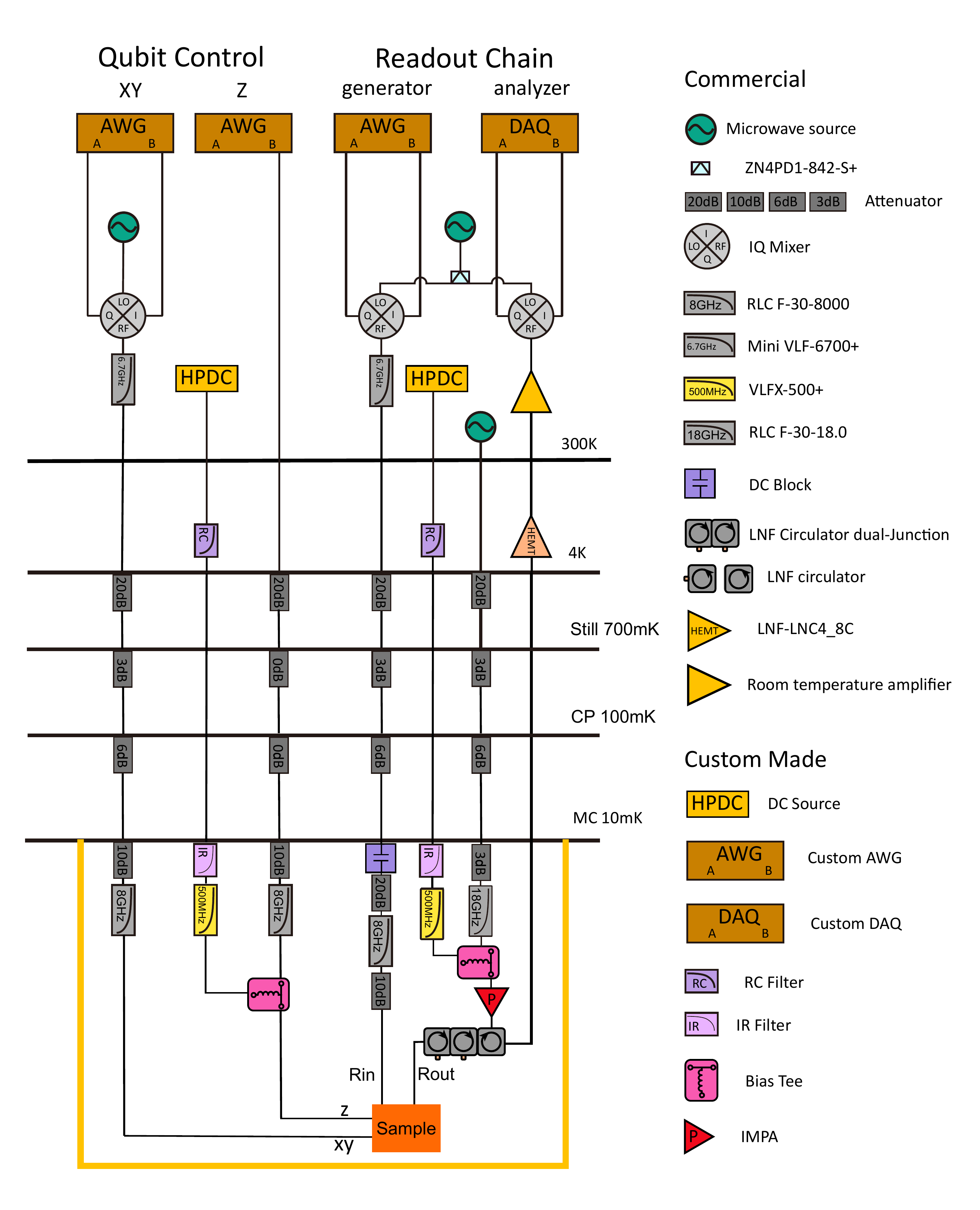}
\caption{Schematic diagram of control electronics and wiring. }
\label{fig:figureS1}
\end{figure}

\section{Supplementary Note 3: performance comparison of different reset protocols}

\begin{figure}[htb]
\centering
\includegraphics[width=0.8\linewidth]{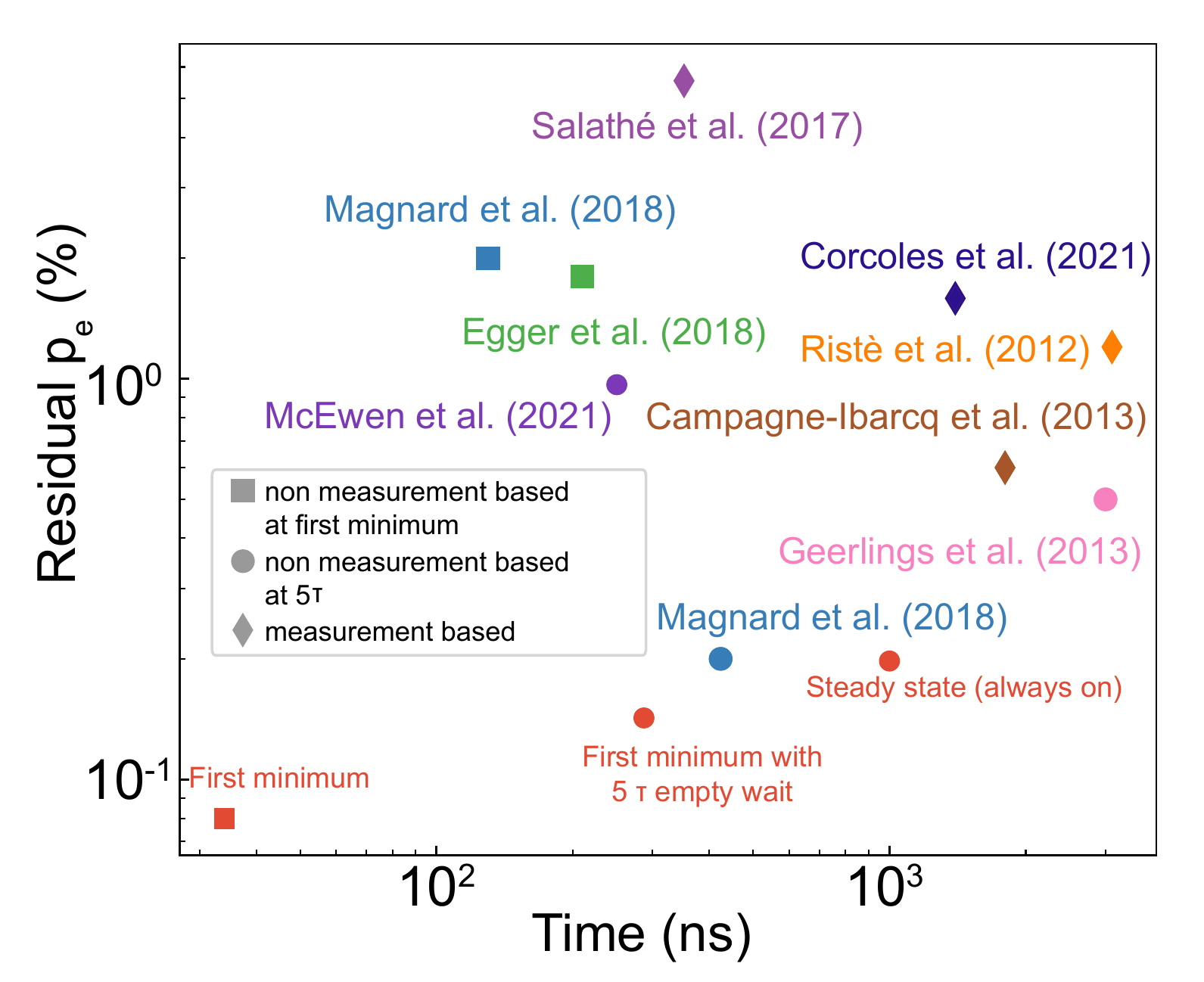}
\caption{Performance comparison of different measurement-based \cite{salathe2018,riste2012,campagne2013,corcoles2021exploiting} and non-measurement-based~\cite{magnard2018,egger2018,geerlings2013,mcewen2021removing} reset protocols. Measurement-based protocols are represented by rhombus. Non-measurement based protocols are represented by squares (first minimum) or circle ( 5$\tau = 5/\kappa_r$ after the first minimum. The main difference between our two protocols denoted by red circles is whether or not the parametric drive is always on during the reset time. }
\label{fig:figureS2}
\end{figure}

\begin{table}[htb]
\centering
\includegraphics[width=1.0\linewidth]{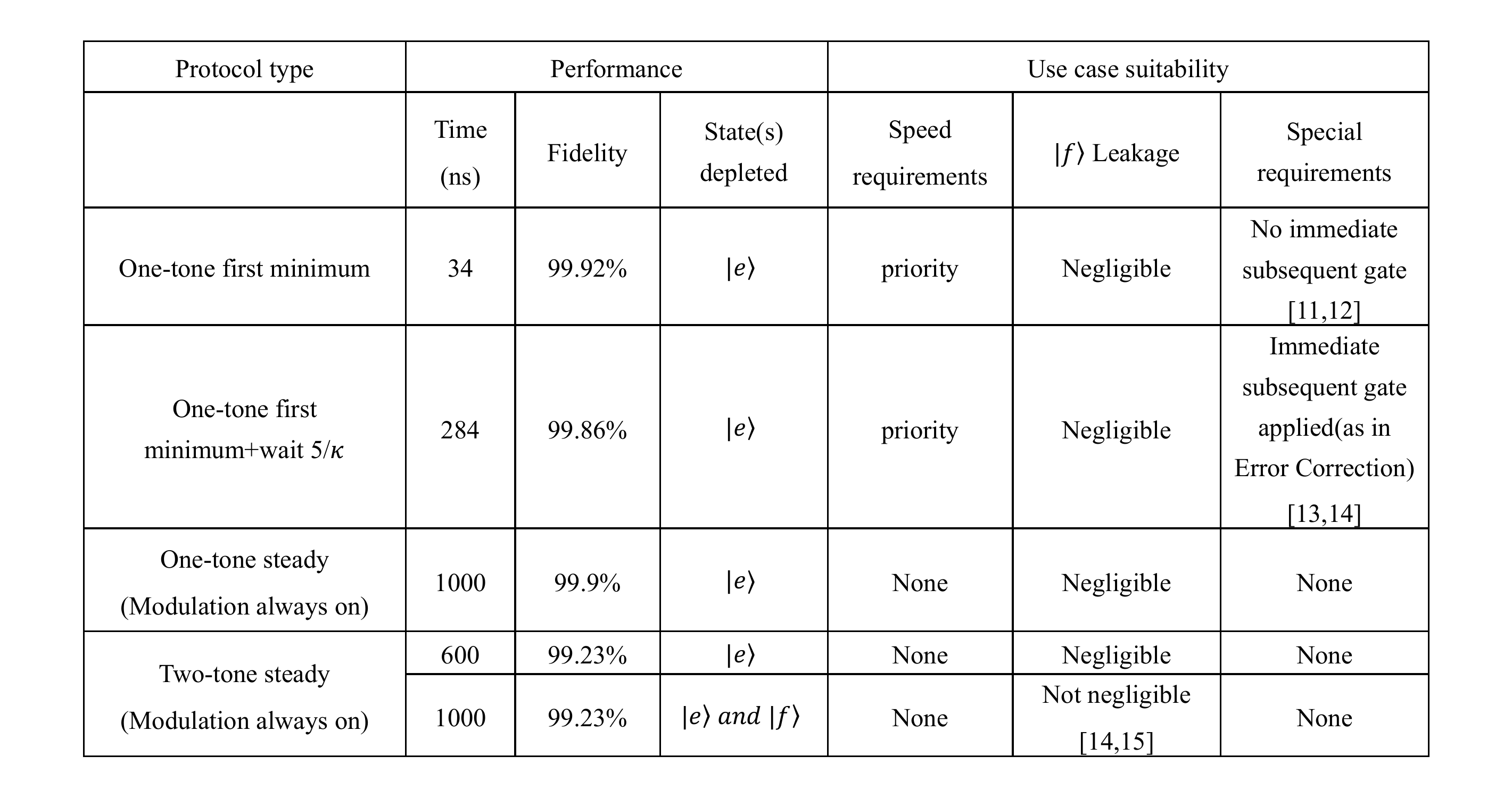}
\caption{Performance and use-case suitability~\cite{kurpiers2018, ma2019,fowler2012surface,marques2021logical,SankPRL16} for our protocols.
}
\label{tab:params2}

\end{table}

Supplementary Fig.~\ref{fig:figureS2} compares various reset protocols for superconducting qubits in terms of two critical parameters: time (speed) and residual $\ket{e}$ state population (fidelity). Two main types of schemes are included: measurement and non measurement-based protocols. In non measurement-based protocols, we summarize the performance at both the first minimum point, and at the point an additional extra 5$\tau = 5/\kappa_r$ after the first minimum (to allow the resonator time to decay to the ground state). In the measurement-based protocols, the time reported is the total duration of the first measurement pulse, the conditional $\pi$ pulse, and the time interval between the two. The performance and use-case suitability of all the protocols we proposed is summarized in Supplementary Table~\ref{tab:params2}.

\section{Supplementary Note 4: Time evolution of excited state population}
\begin{figure}[htb]
\centering
\includegraphics[width=0.9\linewidth]{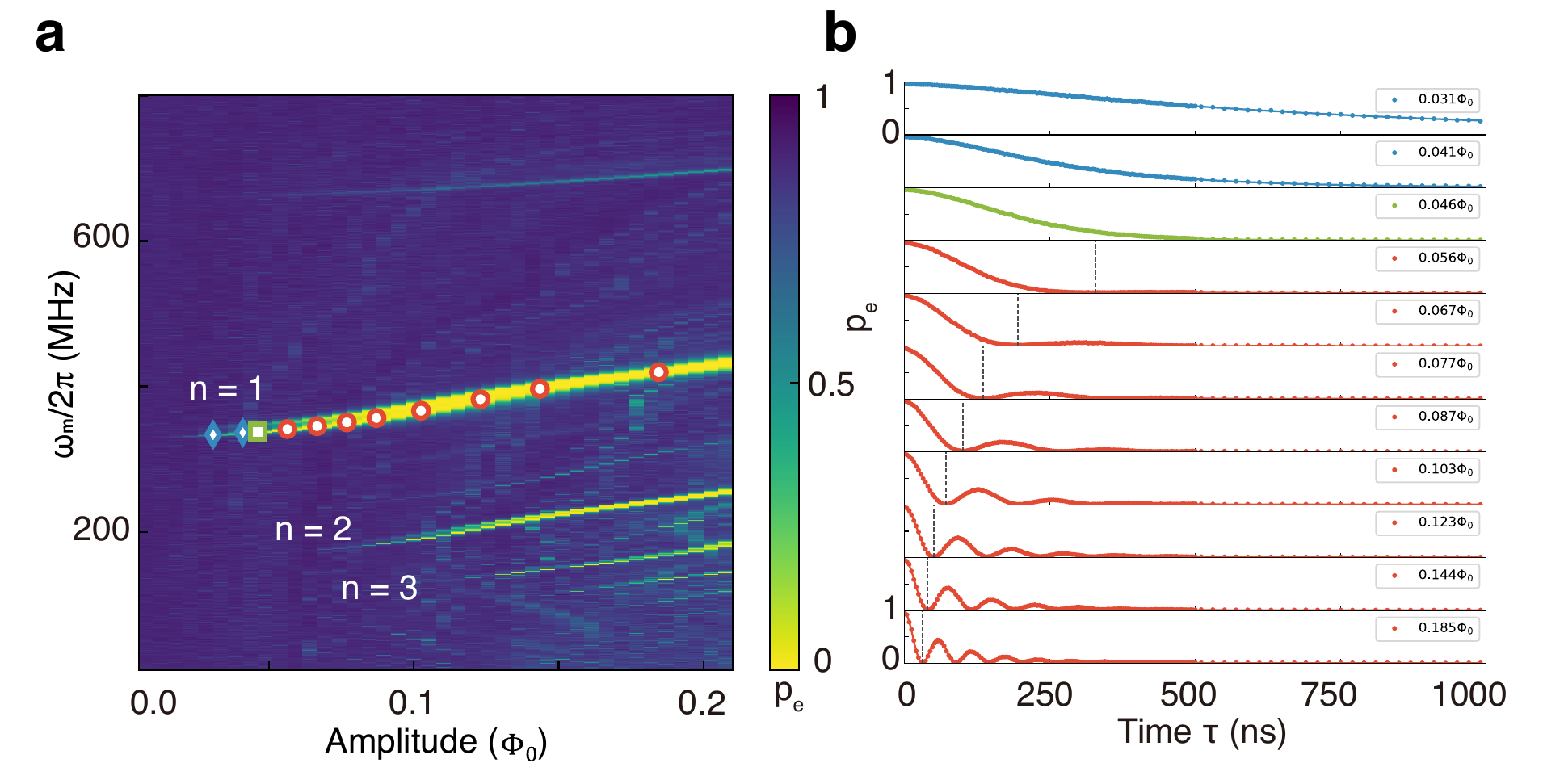}
\caption{Time evolution of different parameters during the parametric reset process. \textbf{a} Two dimensional scan map of $\ket{e}$ population, $p_e$. \textbf{b}  Population $p_e$ of the excited state $\ket{e}$ as a function of reset duration $\tau$, for the points shown in \textbf{a}. Red (blue) symbols represent the underdamped (overdamped) regime, and green symbols correspond to parameters close to the critically damped point. The black dashed lines indicate the first minimum of $p_e$. Modulation amplitudes are shown in the upper right corner of each panel.}
\label{fig:figureS3}
\end{figure}

The population of the resonator-qubit state $\ket{s}$ during the parametric reset process
can be expressed as
\begin{equation}\label{eq:popt_general}
P_{s|s_0}(t) = |\bra{s}\exp(-i H_\mathrm{eff} t)\ket{s_0}|^2.
\end{equation}
where $\ket{s_0}$ is the initial state of the system. When the system is prepared in $\ket{e,0}$, 
the population of the qubit excited state $p_e$ can be shown to be
\begin{equation}\label{eq:pe}
p_e = P_{e|e}(t)= \begin{cases}
e^{-\frac{\kappa_r t}{2}} \left(\frac{\kappa_r t}{4} + 1\right)^2 & |g_n| = \kappa_r/4\\
e^{-\frac{\kappa_r t}{2}} \left[\cos(Mt)+\frac{\kappa_r}{4M}\sin(Mt)\right]^2
& |g_n|>\kappa_r/4,\, M=\frac{\sqrt{16|g_n|^2-\kappa_r^2}}{4}\\
e^{-\frac{\kappa_r t}{2}} \left[\cosh(Mt)+\frac{\kappa_r}{4M}\sinh(Mt)\right]^2
& |g_n|<\kappa_r/4,\, M=\frac{\sqrt{\kappa_r^2 - 16|g_n|^2}}{4}\\
\end{cases}
\end{equation}
When fitting to experimental data, we use the function
\begin{equation}\label{eq:fit_func}
f(t) = \lambda p_e + \mu
\end{equation}
where $\lambda$ and $\mu$ are introduced to account for SPAM errors.  

Supplementary Fig.~\ref{fig:figureS3}a displays the same two-dimensional scan data as Fig.~2b of the main text. At each of the amplitude-frequency parameter pairs corresponding to the colored markers in the $n=1$ strip, we measure the excited population against $p_e$ as a function of the reset duration $\tau$, with results given in Supplementary Fig.~\ref{fig:figureS3}b.
From top down, the modulation amplitude increases from $0.03\Phi_0$ to $0.185\Phi_0$. Colored dots in the traces are experimental results, and solid lines are fittings to the theoretical model (Supplementary Equation ~(\ref{eq:fit_func})).  
We use the same $\lambda$ and $\mu$ across the traces, as these parameters are expected to be independent of modulation amplitude.

The traces in blue shows a decay without any observable oscillation.  From the fitting, we verify that $|g_n|/4 < \kappa_r$  ($|g_n|/\kappa_r\approx0.127, 0.195$ respectively), corresponding to the overdamped regime. 
The population decays faster when the modulation amplitude increases, which agrees with the theoretical prediction (equation (4) in the main text).
For the population in green, no oscillation is observed, and the parameter $|g_n|/4\kappa_r$ is approximately 0.268, very close to the crictical value ($|g_n|/4\kappa_r=0.25$). 
The red traces are in the underdamped regime where oscillations can be observed. The dashed black lines in these panels represent the times when the population achieves a minimum for the first time. In the underdamped regime, we see that Rabi oscillations become more pronounced as the modulation amplitude increases, and the time to the first minimum decreases.
By fitting to the theoretical model (Supplementary Equation~(\ref{eq:fit_func})),
we find that the parameter $|g_n|/4\kappa_r$ increases from 0.421 to
2.195 as the modulation amplitude increases from $0.056\Phi_0$ to $0.185\Phi_0$.

\section{Supplementary Note 5: Q1 operation point away from the sweet spot}
When the operation point for qubit Q1 is away from the sweet spot, the $\ket{e}$ state population is monitored by varying the modulation frequency $\omega$ and amplitude $A$, as displayed in Supplementary Fig.~\ref{fig:figureS4}. Several strip-shaped regions labeled $n = 1,2,3$ can be seen which correspond to the $n$-th order modulations, in which parametric resets can be performed. There are also several strip-shaped lines in the bottom left corner, which we attribute to parametric-modulation-induced qubit-qubit interactions with qubits Q2 and Q3. Supporting this assumption is the observation that, with other scan parameters unchanged, if Q2 or Q3 is tuned far from their sweet spots using DC bias, the corresponding strips disappear.
\begin{figure}
\centering
\includegraphics[width=0.6\linewidth]{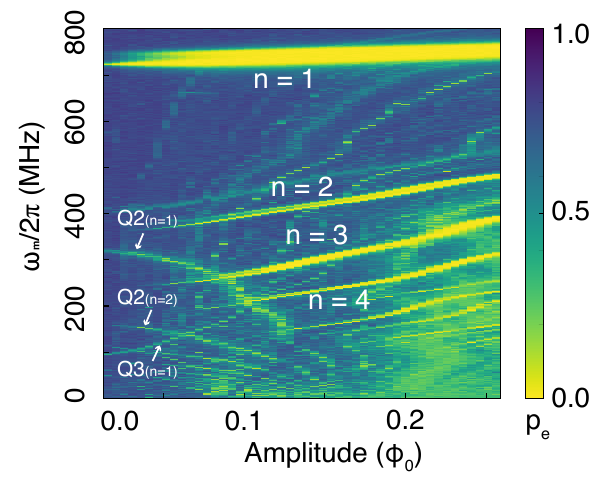}
\caption{Q1 operation point away from sweet spot. Excited state population of Q1 as a function of parametric reset amplitude and frequency.}
\label{fig:figureS4}
\end{figure}

\section{Supplementary Note 6: Residual excited population limit estimation and the thermal occupation of resonator}

\begin{figure}
\centering
\includegraphics[width=0.8\linewidth]{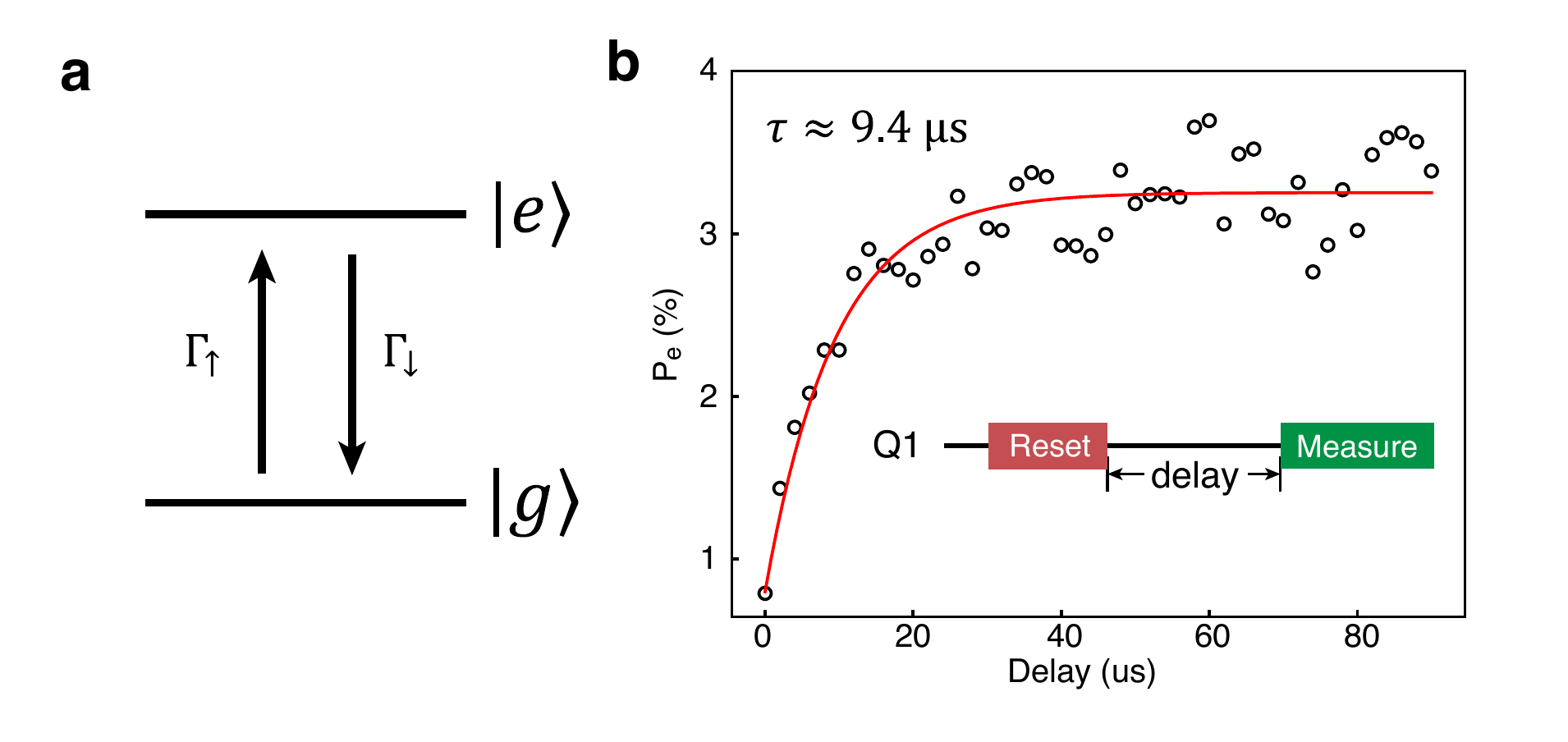}
\caption{\textbf{a} Rate equation model with excited rate $\Gamma_{\uparrow}$ and qubit reset rate $\Gamma_{\downarrow}$. \textbf{b} Re-thermalization after reset.}
\label{fig:figureS5}
\end{figure}
Supplementary Fig.~\ref{fig:figureS5}a illustrates the rate equation model for a qubit with ground and excited states $\ket{g}$ and $\ket{e}$, and excitation and reset rates $\Gamma_{\uparrow}$ and $\Gamma_{\downarrow}$. When the system reaches thermal equilibrium, the final residual $\ket{e}$ population $p_e$ is given by $p_e = \Gamma_{\uparrow}/(\Gamma_{\uparrow}+\Gamma_{\downarrow})$. Before reset, 
$p_e$ is measured to be around 2.38\% from Rabi population
measurement (RPM) as described in the main text. From the measured $T_1$ we deduce that $1/T_1 = \Gamma_{\uparrow}+\Gamma_{\downarrow} = 86.6 \mathrm{kHz}$. When applying parametric reset, $\Gamma_{\uparrow}$ remains the same value while $\Gamma_{\uparrow}+\Gamma_{\downarrow}$ is 10 MHz in the underdamped region ($\kappa_{r}/2$, the exponential term in Supplementary Equation~(\ref{eq:pe}) when $|g_n|>\kappa_r/4$).
Accordingly, $p_{e\mathrm{(reset)}}$ drops to $0.02\%$, a rough theoretical estimate of the residual excitation population limit. 
After the reset, the system will be re-thermalized to the initial steady state as shown in Supplementary Fig.~\ref{fig:figureS5}b. 
The raw data is fit to a single exponential function with time constant 9.4 $\mu$s.

When the system begins with $\ket{g}$, there is no experimental observation of $\ket{e}$ or $\ket{f}$ population. Based on our state readout fidelity, we conclude that the observed excitation probability below 0.1\%. The total number of excitations should be conserved during the reset protocol, and the reset frequency (hundreds of MHz) is much smaller than the qubit and resonator energy gap. The likelihood of re-excitation during the reset modulation should thus be negligible.

When the average thermal population of the resonator $\overline{n} \ll 1$, the thermal population induced dephasing is propotional to $\overline{n}$ \cite{clerk2007using}, i.e.

\begin{equation}
\Gamma_{\phi} = \frac{1}{T_2^*}=\frac{\overline{n}\kappa\chi^{2}}{\chi^{2}+\kappa^{2}}
\end{equation}
where $\kappa$ is the resonator linewidth and $\chi = g_{qr}^{2}/(\Delta(1+\Delta/\eta))$ is the dispersive shift. From the measured value of $T_2^*,g_{qr},\eta,\kappa,\Delta =\omega_{r}-\omega_{q}$ in Supplementary Table~\ref{tab:params} and main text, and assuming that thermal population is the only source of dephasing, we deduce  $\overline{n} \leq 0.01$ (corresponding to a negligible stark shift of 51.5 kHz). 
This estimate gives an upper bound on the thermal population.

\section{Supplementary Note 7: Ramsey experiments to study the effects of the parametric drive on neighbouring qubits}
To investigate the effect of the parametric modulation pulse on neighbouring qubits,  we perform a series of Ramsey measurement on one qubit. During the interval $\tau$ -- ranging from 100 to 1000 ns -- between two $\pi/2$-pulses, on another qubit(s) we either apply parametric modulation or, for comparison, do nothing. The pulse sequence is presented in Supplementary Fig.~\ref{fig:figures6}a. The amplitude and phase of each Ramsey fringe can be extracted by sinusoidal fitting. We investigate the following cases: Supplementary Fig.~\ref{fig:figures6}b,c: Ramsey on Q2 while Q1 and Q3 are reset simultaneously. Supplementary Fig.~\ref{fig:figures6}d,e: Reset on Q2 and Ramsey on Q1 or Q3 respectively. Supplementary Fig.~\ref{fig:figures6}f,g: Reset on Q1 and Ramsey on Q2 or Q3, respectively. In all cases, we observe that the Ramsey amplitude decreases with $\tau$, which is reasonable due to inevitable dephasing. More importantly, 
no additional degradation of the coherence is observed when the parametric
reset modulation is applied on neighbouring qubits. With $\tau$ increasing from 100 ns
to 1000 ns, the phase remains steady with fluctuations below 1 rad. 
From a linear fitting to each case, we determine that the frequency 
shift due to the parametric modulation pulse is in the tens of kHz. These experiments demonstrate that our parametric reset protocol has
negligible effects on neighbouring qubits in terms of both coherence and frequency (crosstalk).

\begin{figure}
\centering
\includegraphics[width=0.8\linewidth]{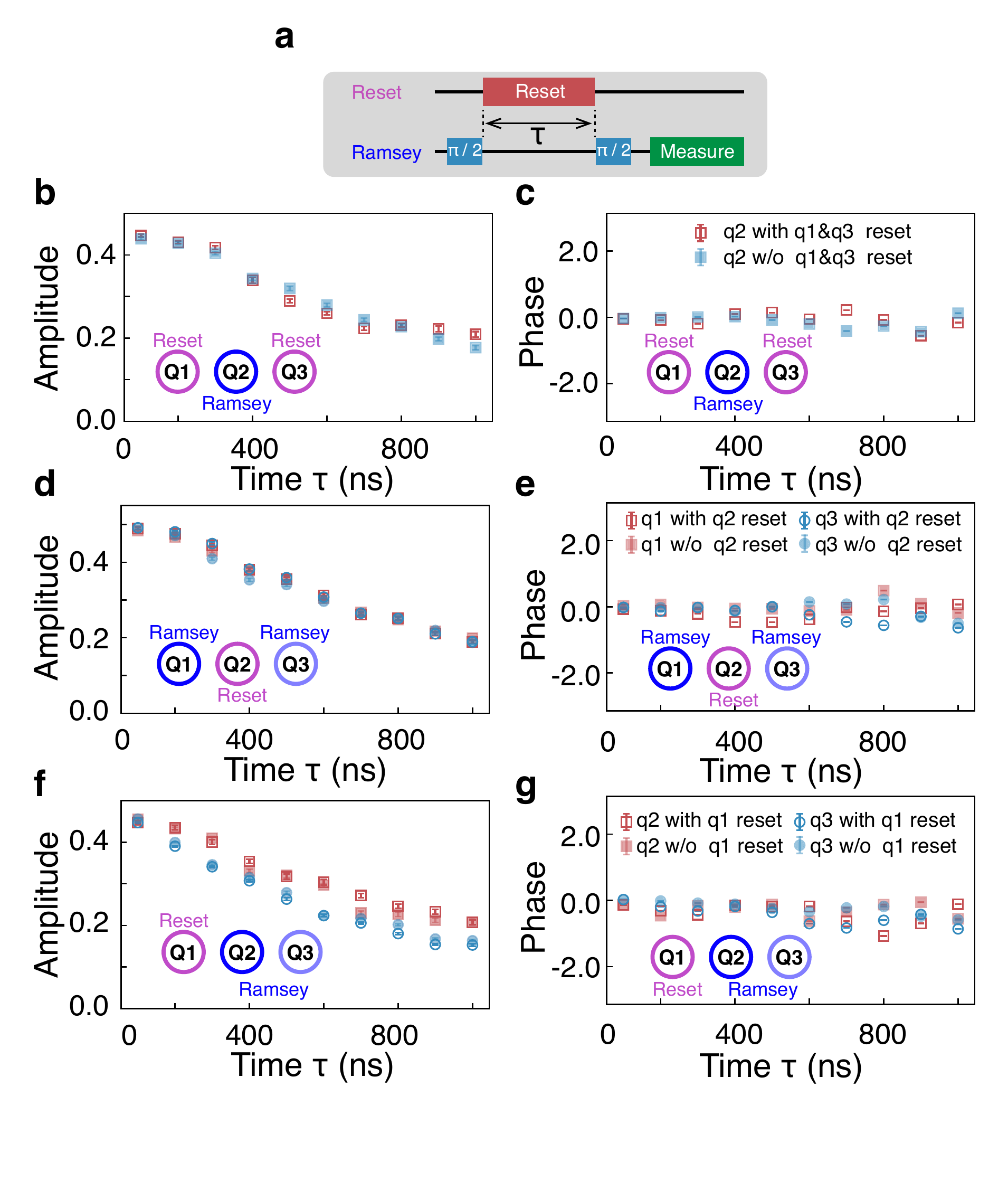}
\caption{Ramsey based experiments. \textbf{a} Pulse sequence: Ramsey measurement on one qubit by changing the phase of the second $\pi/2$-pulse. During the interval $\tau$ between two $\pi/2$-pulses, the other qubit is reset. The Ramsey amplitude and phase are extracted by fitting sinusoids to the Ramsey fringes at each $\tau$. \textbf{b,c} Ramsey on Q2 while Q1 and Q3 are reset simultaneously.\textbf{d,e} Reset on Q2 and Ramsey on Q1 or Q3, respectively. \textbf{f,g} Reset on Q1 and Ramsey on Q2 or Q3, respectively. All error bars are statistical ($\pm$1 s.d.). No additional effects of decoherence or frequency shifts are observed.}
\label{fig:figures6}
\end{figure}

\section{Supplementary Note 8: Reset of the $|f\rangle$ state}

Here, we briefly describe the theory behind the two-tone reset protocol used in the main text to reset both the $\ket{e}$ and $\ket{f}$ states. The Hamiltonian of the qubit-resonator system can be expressed as:
\begin{equation}
H_S = \sum_{n} (n\omega_q(t) + \eta_{n})\ketbra{n}_q + \omega_r a^\dagger a + g_{qr}\sum_{n} \sqrt{n+1}( \ketbra{n}{n+1}_q a^\dagger + \ketbra{n+1}{n}_q a)
\end{equation}
where $\omega_q(t)$ is the time dependent qubit frequency under modulation, $a$ is the resonator annihilation operator, $g_{qr}$ is the coupling strength between qubit and resonator, and $\eta_n$ is the anharmonicity of the qubit, $\eta_{0,1}=0, \eta_2 = \eta$.
We consider the three lowest levels of the resonator, and during the modulation of the qubit, the frequency $\omega_q(t)$ can be expanded as
\begin{equation}\label{eq:freq_mod_2tone}
\omega_q(t) = \overline{\omega_q} + \sum_{k=1}^M A_k \cos\omega_k t
\end{equation}
In the interaction picture, the Hamiltonian reads:
\begin{equation}
H_\mathrm{int}(t) = g_{qr} \exp(i\sum_{k=1}^M \frac{A_k}{\omega_k} \sin\omega_k t)\left[e^{i\overline{\Delta}t}\ketbra{e0}{g1}+e^{i(\overline{\Delta}+\eta)t}\ketbra{f0}{e1}+e^{i\overline{\Delta}t}\ketbra{e1}{g2} + h.c. \right]
\end{equation}
where $\overline{\Delta}=\overline{\omega_q}-\omega_r$
Expanding the exponential term $\exp(i\sum_{k=1}^M\frac{A_k}{\omega_k} \sin(\omega_k t))$ using the identity $e^{iy \sin x}=\sum_{n\in \mathbb{Z}} J_n(y) e^{inx}$, we obtain:
\begin{equation}
\begin{aligned}
\exp(i\sum_{k=1}^M\frac{A_k}{\omega_k} \sin\omega_kt) 
&= \prod_{k=1}^M
J_0\left(\frac{A_k}{\omega_k}\right) \\ 
&\quad + \sum_{k=1}^M J_1\left(\frac{A_k}{\omega_k}\right)(e^{i\omega_k t} - e^{-i\omega_k t}) \left(\prod_{m\neq k} J_0\left(\frac{A_m}{\omega_m}\right)\right) \\
&\quad + \ldots
\end{aligned}
\end{equation}
By ignoring the higher order terms, the Hamiltonian in the interaction picture becomes:
\begin{equation}\label{eq:ham_2tone}
H_\mathrm{int}(t) \approx \left[g_0 + \sum_{k=1} g_{1,k}(e^{i\omega_k t} - e^{-i\omega_k t})\right]\left[e^{i\overline{\Delta}t}\ketbra{e0}{g1}+e^{i(\overline{\Delta}+\eta)t}\ketbra{f0}{e1}+e^{i\overline{\Delta}t}\ketbra{e1}{g2} + h.c. \right]
\end{equation}
where
\begin{equation}
\begin{aligned}
g_0 &= g\prod_{k=1}^M J_0\left(\frac{A_k}{\omega_k}\right)\\
g_{1,k} &= g J_1\left(\frac{A_k}{\omega_k}\right) \left(\prod_{m\neq k} J_0\left(\frac{A_m}{\omega_m}\right)\right)
\end{aligned}
\end{equation}
From Supplementary Equation~(\ref{eq:ham_2tone}), when one of the $\omega_k$ satisfies $\omega_k = \pm \overline{\Delta}$,
a swap between $\ket{e0}$ and $\ket{g1}$ occurs, and the qubit population of $\ket{e}$ can be reset through the resonator.
Similarly, when one of the $\omega_k$ satifies $\omega_k = \pm (\overline{\Delta}+\eta)$, a swap between $\ket{f0}$ and $\ket{e1}$ is activated, which enables the reset of the $\ket{f}$.
As the qubit frequency is modulated with multiple frequencies (Supplementary Equation~(\ref{eq:freq_mod_2tone})), the reset of $\ket{e}$ and $\ket{f}$ can be achieved simultaneously through different frequencies.

In the main text, the flux is modulated by two frequencies $\omega_{1,2}$, and the qubit frequency $\omega_q(t)$ has four main Fourier components $2\omega_1, 2\omega_2, \omega_1\pm\omega_2$. 
Thus $2\omega_{1,2}=-\Delta$ or $\omega_1+\omega_2 = -\Delta$ enables the reset of $\ket{e}$, and $2\omega_{1,2}=-\Delta-\eta$ or $\omega_1+\omega_2 = -\Delta - \eta$ enables the reset of $\ket{f}$.
When $\omega_1 = -\Delta/2$ and $\omega_2 = -\Delta/2-\eta$, both $\ket{e}$ and $\ket{f}$ are reset (rhombus R in Fig.4b in the main text)

The full scan-maps of our two-tone parametric reset are displayed in Supplementary Fig.~\ref{fig:figures7} with 1000 ns reset time and qubit prepared in the $\ket{f}$ state before reset (c.f. Fig.~4b and c in the main text). 
The first and second rows are experimental and master equation simulation data, respectively, with good agreement between the two. 
The yellow strips in the $P_f$ scan-map are located in the same positions as the blue strips in the $P_e$ scan-map, which indicates that a swap between $\ket{f0}$ and $\ket{e1}$ -- and thus the depletion of $\ket{f}$ -- occurs in this region. 
Similarly, the relative positions of the yellow strips in the $P_e$ map and the blue strips in the $P_g$ map indicate an exchange between $\ket{e0}$ and $\ket{g1}$, corresponding to a reset of the $\ket{e}$ state.
Small deviations between experimental and simulated data can be observed in Supplementary Fig.~\ref{fig:figures7}, which we attribute to a mismatch between the transfer functions of the z pulse in simulation and experiment.

\begin{figure}
\centering
\includegraphics[width=1.0\linewidth]{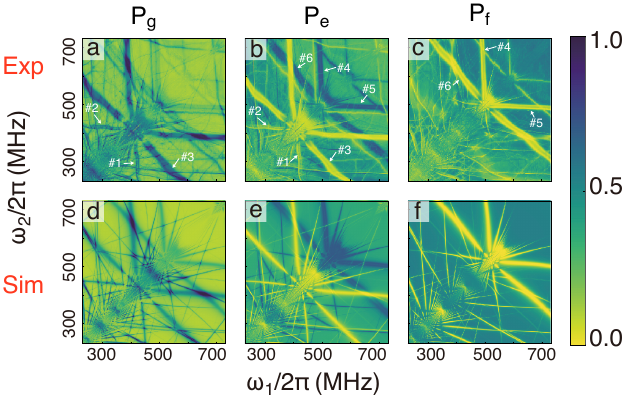}
\caption{two-tone parametric reset \textbf{a-c} The qubit is prepared in the $\ket{f}$ state and then reset with 1000 ns two-tone parametric reset pulse, after the reset, the population of  $P_g$,$P_e$,$P_f$ is measured in \textbf{a}, \textbf{b}, \textbf{c} respectively. \textbf{d-f} Master equation simulation of the whole process with the same parameters as in \textbf{a-c}}
\label{fig:figures7}
\end{figure}

\section{Supplementary Note 9: Repeated reset}
The data in Supplementary Fig.~\ref{fig:figures8} verifies that our reset protocols can be consistently reapplied without accumulation of errors. In \textbf{a}, the qubit is repeatedly prepared in the $\ket{e}$ state and then reset via single-tone modulating pulse. In \textbf{b}, the qubit is repeatedly prepared in the $\ket{f}$ and reset via a two-tone modulating pulse. In both figures, the excitation-reset procedure is repeated up to 100 times. No significant errors are observed during these repeated reset processes.

\begin{figure}
\centering
\includegraphics[width=0.8\linewidth]{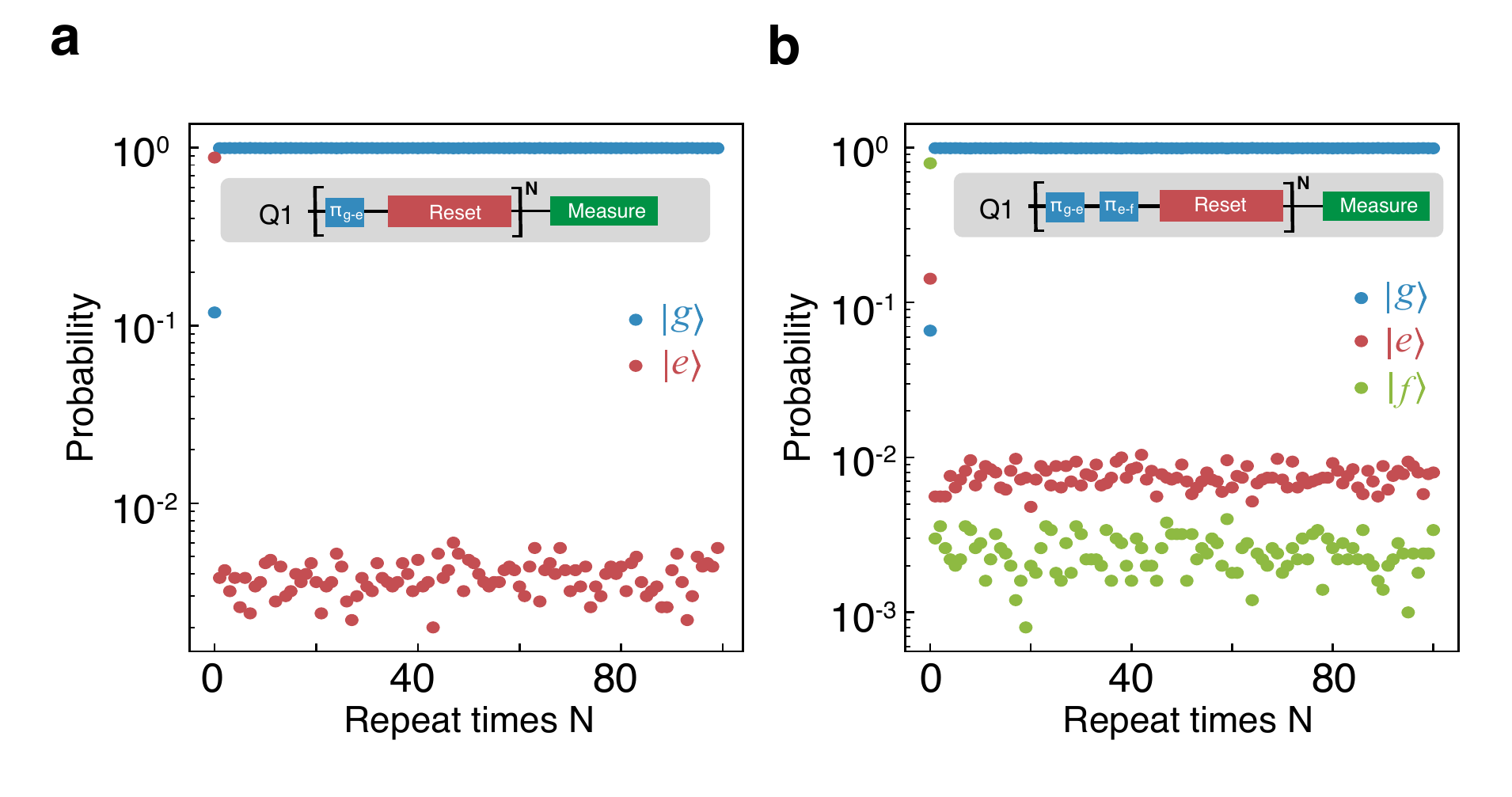}
\caption{Repeated reset. \textbf{a} Initial $\ket{e}$ state  with single-tone parametric reset repeated for 100 times. \textbf{b} Initial $\ket{f}$ state  with two-tone parametric reset repeated for 100 times.}
\label{fig:figures8}
\end{figure}

\section{Supplementary Note 10: RB of the two-tone parametric reset}
The two-tone parametric reset pulse has more frequency components than the one-tone pulse. Those unused frequency components may introduce gate errors in neighbouring qubits. We study these effects in Supplementary Figure ~\ref{fig:figures9}. During the two-tone reset of Q1,  Clifford randomized benchmarking (RB) is performed on the nearest-neighbour (Q2) or next-nearest-neighbour (Q3) qubit. For comparison, we perform RB on Q2 and Q3 without any rest on Q1. The error bars are statistical ($\pm$1 s.d.) with 30 repetitions. We find the average gate fidelity variation between whether the reset is applied or not is 0.06$\%$ (0.01$\%$) on Q2 (Q3).

\begin{figure}
\centering
\includegraphics[width=0.8\linewidth]{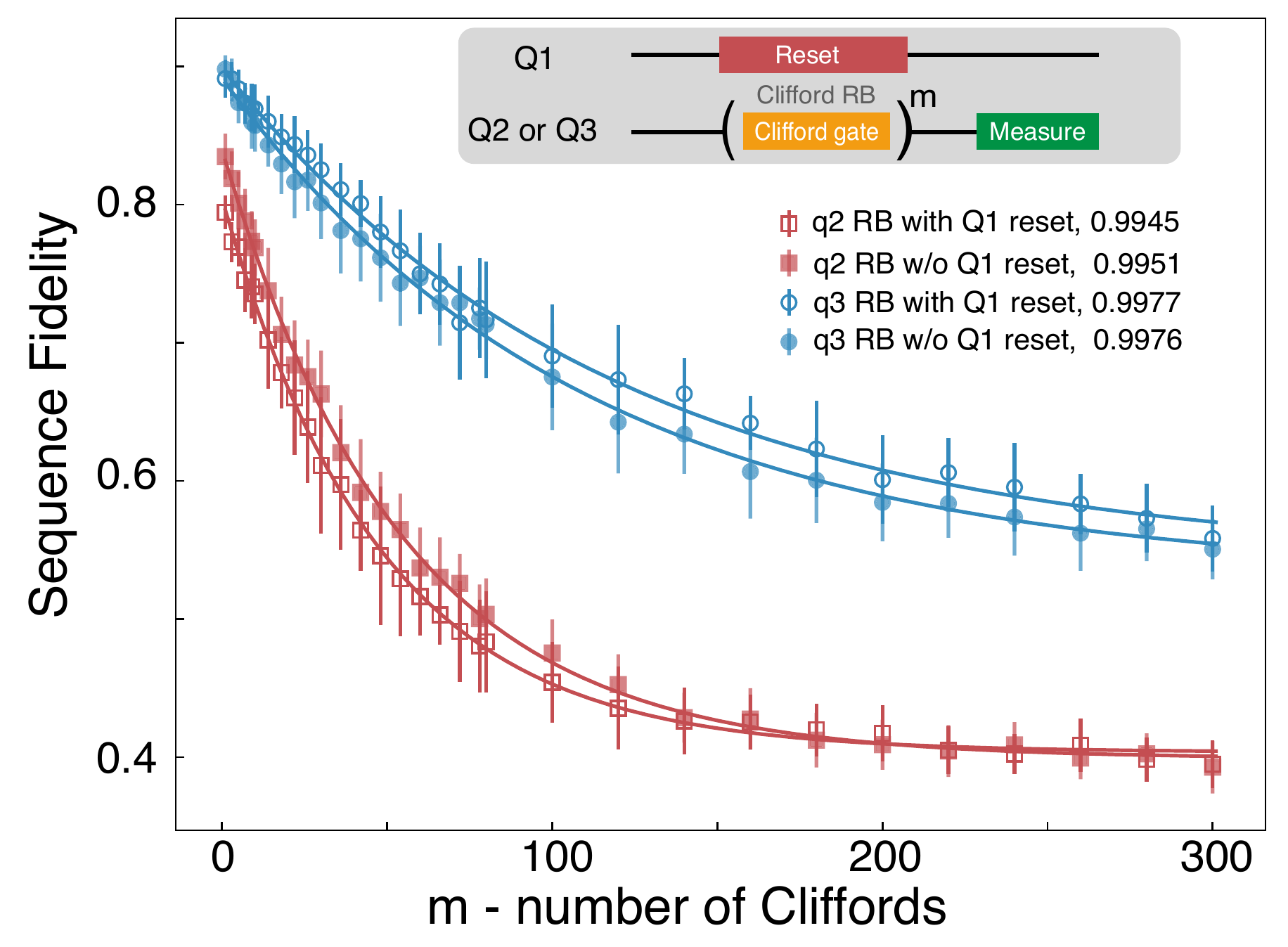}
\caption{Clifford RB on neighbouring qubits. \textbf{a} Initial $\ket{e}$ state  with single-frequency reset, repeated 100 times. \textbf{b} Initial $\ket{f}$ state  with two-frequency reset, repeated 100 times. The error bars are statistical ($\pm$1 s.d.) with 30 repetitions.}
\label{fig:figures9}
\end{figure}

\section{Supplementary Note 11: Ramsey fringes fitting with the beating pattern}
We observed beating patterns in the Ramsey fringes of Q1 and Q3, as shown in Supplementary Figure~\ref{fig:figures10} a and c. We fitted those data by

\begin{equation}
\begin{aligned}
P_t = e^{-t/T_2} (A_1\cos({\omega_1} t+\phi_1)+A_2\cos(\omega_2 t+\phi_1))+B
\end{aligned}
\end{equation}
as in \cite{peterer2015coherence}. After removing the $T_1$ contribution in the fitted $T_2$, we extracted pure dephasing $T_{2}^{*}$  in Supplementary Table 1 (for Ramsey fringes of Q2, no beating pattern is observed, only one oscillation term is used in the fitting). The beating is usually related to the noise in the environment. We performed spin-echo measurements of Q1 and Q3 ,and the beating patterns disappear, suggesting that the noise can be refocused by adding a $\pi$ pulse between two $\pi/2$ pulses. 

\begin{figure}
\centering
\includegraphics[width=0.85\linewidth]{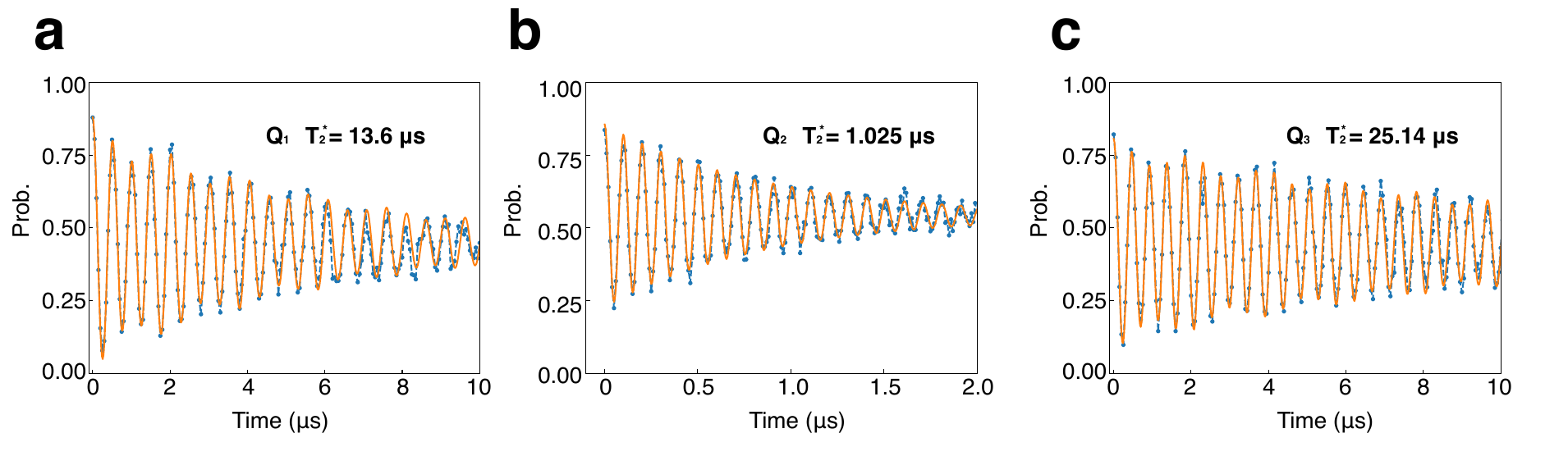}
\caption{Ramsey fringes of the three qubits.}
\label{fig:figures10}
\end{figure}
